\documentclass[a4paper,11pt]{article}
\usepackage{amsfonts,amsthm,amsmath,amssymb,graphicx,hyperref,color}

\usepackage[vcentermath]{youngtab}

\def\bra#1{\langle #1 |}
\def\ket#1{| #1\rangle}
\def\mbra#1{\{ #1 |}
\def\mket#1{| #1\}}
\newcommand{\braket}[2]{\langle #1 | #2 \rangle}
\newcommand{\mbraket}[2]{\{ #1 | #2 \}}

\def\corr#1{\left\langle #1 \right\rangle}
\newcommand{\tr}{\operatorname{tr}}

\newcommand{\mult}{\operatorname{mult}}

\newcommand{\nat}{\operatorname{nat}}
 
\newcommand{\Sym}{\operatorname{Sym}}

\def\Dim{\textrm{Dim}}

\def\Sym{\textrm{Sym}}

\def\a{\alpha}
\def\b{\beta}

\def\g{\gamma}

\def\e{\epsilon}

\def\m{\mu}

\def\n{\nu}

\def\O{\Omega}
\def\r{\rho}
\def\k{\kappa}

\def\s{\sigma}

\def\S{\Sigma}
\def\l{\lambda}
\def\L{\Lambda}

\def\d{\partial}

\def\D{\Delta}

\def\C{\mathbb{C}}

\def\R{\mathbb{R}}

\def\Z{\mathbb{Z}}
\def\Q{\mathbb{Q}}

\newcommand{\cN}{\mathcal N}
\newcommand{\cO}{\mathcal O}

\newcommand{\be}{\begin{equation}}
\newcommand{\bea}{\begin{eqnarray}}

\newcommand{\ee}{\end{equation}}
\newcommand{\eea}{\end{eqnarray}}

\newcommand{\nn}{\nonumber}

\def\tyng(#1){\hbox{\tiny$\yng(#1)$}\,}
\def\tyoung(#1){\hbox{\tiny$\young(#1)$}}

\newcommand{\pd}{\!\!+\!\!}

\title{Cut-and-join operators and $\cN=4$ super Yang-Mills}

\begin{document}

\rightline{DESY 10-020}

\vspace{4truecm}

\centerline{\LARGE \bf Cut-and-join operators and $\cN=4$ super Yang-Mills}

\vspace{1truecm}

\centerline{{\large \bf T.W. Brown${}^\star$}}

\vspace{.4cm}
\centerline{{\it  DESY Hamburg, Theory Group,}}
\centerline{{\it  Notkestrasse 85, D-22607 Hamburg, Germany.}}

\vspace{1.5truecm}

\thispagestyle{empty}

\centerline{\bf ABSTRACT}

\vspace{.5truecm}

\noindent We show which multi-trace structures are compatible with the
symmetrisation of local operators in $\cN=4$ super Yang-Mills when
they are organised into representations of the global symmetry group.
Cut-and-join operators give the non-planar expansion of correlation
functions of these operators in the free theory.  Using these
techniques we find the  $1/N$ corrections to the quarter-BPS
operators which remain protected at weak coupling.  We also present a
new way of counting these chiral ring operators using the Weyl group
$S_N$.

\vfill
\noindent{\it ${}^\star$thomas.william.brown@desy.de}

\newpage

\tableofcontents

\setcounter{footnote}{0}

\section{Introduction and summary}

Recent work on solving $\cN=4$ super Yang-Mills with gauge group
$U(N)$ for finite $N$ at weak coupling has focused on a basis of local
operators which is well-defined regardless of the number of fields the
operators contain \cite{0111222}-\cite{0910.2170}.  In particular for
classical conformal dimensions $\D \geq N$ the operators accommodate
the Stringy Exclusion Principle \cite{9804085} which limits the
operators constructible from finite-size matrices.  All the bases have
been defined by generalising the half-BPS Schur polynomials
\cite{0111222} to the case where multiple fields of the theory are
included and sorted into representations of the global symmetry group
$PSU(2,2|4)$ or its subgroups.  Like the Schur polynomials these bases
of operators are fully diagonal in the non-planar two-point function
at tree level \cite{0711.0176,0806.1911}, but unlike the Schur
polynomials they mix at one loop \cite{0801.2094}  because they are no longer necessarily BPS.

These operators, which are traced with representations of the gauge
group $U(N)$ like a Wilson loop, are linear combinations of operators
with different trace structures.  The goal of this paper is to return
the non-planar analysis to operators which have a fixed trace
structure for $\Delta < N^{\frac{1}{2}}$ but are still organised into
representations of the global symmetry group.  The latter is important
because we expect the bulk spectrum to be organised into representations
of $PSU(2,2|4)$.  The trace structure plays an important part in
defining the operators with a given conformal dimension at weak
coupling and hence identifying the map to string states in the bulk.

\begin{figure}[t]
\begin{center}
\includegraphics[trim=0 0 0 0 ]{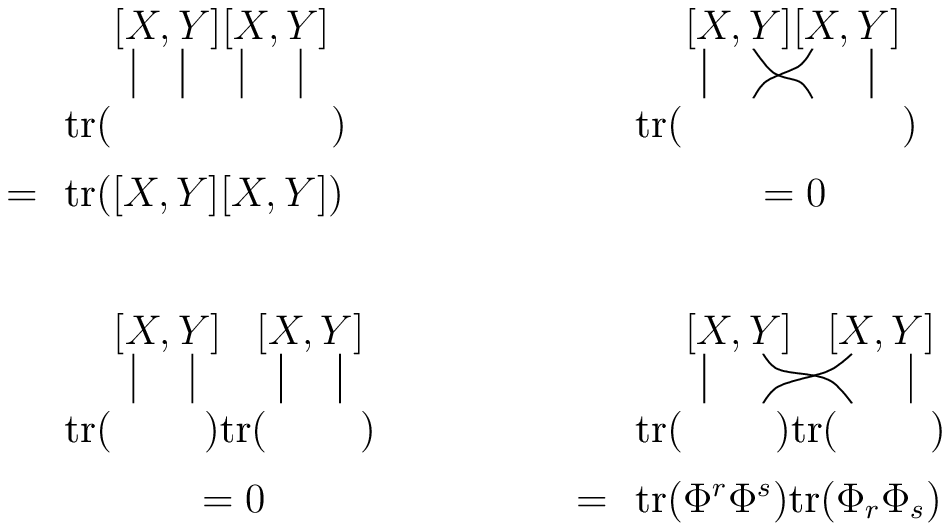}
\caption[]{Different ways of traces the $U(2)$ representation
  $\tyng(2,2)$.}\label{fig:22trace}
\end{center}
\end{figure}

The complications of matching symmetrised representations to traces
can be seen clearly with the $\tyng(2,2)$ representation of the
subgroup $U(2) \subset SU(4)_R \subset PSU(2,2|4)$ of the global
symmetry group. The two columns of the Young diagram correspond to two
commutators of $X$ and $Y$: $[X,Y][X,Y]$.  If we take a single trace
of length 4 of these fields then we get the gauge-invariant operator
$\tr([X,Y][X,Y]) = 2\tr(XYXY) - 2\tr(YXXY)$ (see Figure
\ref{fig:22trace}).  However if we take two traces of length 2 then
the operator $\tr([X,Y])\tr([X,Y])$ vanishes by cyclicity of the trace
since $\tr([X,Y]) = \tr(XY) - \tr(YX) = 0$.  To get a non-trivial
operator with this trace structure you must `twist' the fields with a
permutation before tracing them.  The result is
$\tr(\Phi^r\Phi^s)\tr(\Phi_r\Phi_s)$ where each commutator $[X,Y] =
\Phi^r \Phi_r = \e_{rs} \Phi^r \Phi^s$ crosses between the traces (see
Figure \ref{fig:22trace}).
Also, for a given trace structure and representation there are often
multiple independent operators.

The crucial technical result in this paper is given in Section
\ref{sec:organrep}: for a given trace structure and representation of
the global symmetry group we show how to write down the allowed
operators and count them correctly.  These operators are diagonal in
the planar two-point function. Once we have well-defined operators we
can compute the full non-planar expansion of their free correlation
functions order by order in $\frac{1}{N}$ using the string/spin bit
formalism. For the half-BPS states this is reviewed in Section
\ref{sec:review}.  Every time we cut a trace in two or join two traces
together we gain a factor of $\frac{1}{N}$.  Using cut-and-join
operators in the class algebra of the symmetric group reduces the free
theory combinatorics to multiplications within this algebra
\cite{0209215}.  Mapping the factors of $\frac{1}{N}$ to string
vertices makes the 't Hooft map from non-planar Feynman diagrams to
two-dimensional Riemann surfaces transparent.  Futhermore for large
operators the higher-genus surfaces factorise into planar three-point
functions corresponding to three-punctured spheres in the putative
string dual to free super-Yang-Mills \cite{0205089,0206248,0209215}.
In Section \ref{sec:geninner} we show that this behaviour is not
peculiar to the half-BPS sector but is universal for all operators in
the free theory.  This factorisation gives some evidence that the free
theory becomes simpler when the operators are large.

In the AdS/CFT correspondence \cite{9711200} it is not $\frac{1}{N}$
but rather $g_{s}= g_{\textrm{YM}}^2 = \frac{\l}{N}$ which is
identified with the physical string coupling in the bulk theory.
However, in the free theory both $g_{\textrm{YM}}$ and $\l$ vanish
while $N$ remains finite. In the absence of a definite string theory
dual to the free theory, we identify $\frac{1}{N}$ with the string
coupling in this limit.  It is with respect to this three-string
vertex that the correlation functions factorise.  In the free theory
we also choose to identify single-trace operators with single-string
states and correspondingly for the multi-trace operators, but this
choice is by no means unique \cite{0208231} and does not apply when we
turn on the coupling.

At weak coupling the situation is subtle, but the techniques
developed in Section \ref{sec:general} are still useful to identify the correct states.  In
Section \ref{sec:weak} we consider the $U(2)$ subsector of operators
built from two complex scalars $X,Y$ where the first non-trivial
complications arise.  There is now a separation between those
operators which gain an anomalous dimension when the coupling is
turned on, and those which remain BPS (they are annihilated by a
quarter of the supercharges).  In this rearrangement of operators into
eigenstates of the dilatation operator, different trace structures get
mixed. The main result of this section is an application of the
non-planar technology of Section \ref{sec:general} to find a new way
of writing the $\frac{1}{N}$ corrections for the quarter-BPS operators
based on the earlier work of \cite{0301104} and \cite{0711.0176}.
Much of the free analysis remains, particularly in the fact that
different trace structures come with powers of $\frac{1}{N}$ depending
on how they are cut or joined.

The $U(2)$ operators which remain protected are in one-to-one
correspondence with those operators which can be built assuming that
the fields commute within each trace.  These operators are part of the
chiral ring of the theory.  In Section \ref{sec:weyl} we give a new
way to count the chiral ring by constructing operators which are
functions of the eigenvalues of the fields $X$ and $Y$.

\section{Review of cut-and-join operators for half-BPS operators}\label{sec:review}

In this section we synthesise analysis of the non-planar correlation
functions of multi-trace half-BPS operators from across the literature
\cite{0107119,0111222,0205033,0205089,0209215,0404066}.

The main points of this review are:
\begin{itemize}
\item The Hilbert space of half-BPS operators with $n$ fields maps to
  the conjugacy classes of the symmetric group $S_n$, where the cycles
  of the permutations correspond to traces.
\item Different trace structures are orthogonal in the inner product
  given by the planar two-point function.
\item The $\frac{1}{N}$ corrections to the extremal correlation
  functions of trace operators are captured by the algebra of the
  conjugacy classes of $S_n$.
\end{itemize}

In each half-BPS multiplet there is an operator built from $n$ copies
of the same complex scalar field $X$.  
The trace structure is captured by a permutation $\a \in S_n$
\begin{equation}
  \tr(\a\; X) \equiv X^{i_1}_{i_{\a(1)}} X^{i_2}_{i_{\a(2)}}  \cdots X^{i_n}_{i_{\a(n)}} 
\end{equation}
For example the single-trace operator can be written with a single cycle of length $n$
\begin{equation}
  \tr(X^n) = \tr([n]\; X) = X^{i_1}_{i_{2}} X^{i_2}_{i_{3}}  \cdots X^{i_n}_{i_{1}} 
\end{equation}
In this case $\a$ is an $n$-cycle $\a=[n]\equiv (123\cdots n)$.
Similarly an operator with two traces of lengths $n_1$ and $n_2$ is
written using $\a = [n_1,n_2] = (12\cdots n_1)(n_1+1 \cdots n_1+n_2)$
\begin{equation}
  \tr(X^{n_1})\tr(X^{n_2}) = \tr([n_1,n_2]\; X) = X^{i_1}_{i_{2}} X^{i_2}_{i_{3}} \cdots X^{i_{n_1}}_{i_1} X^{i_{n_1+1}}_{i_{n_1+2}} \cdots X^{i_{n_1+n_2}}_{i_{n_1+1}} 
\end{equation}
The final operator only depends on the conjugacy class of $\a \in S_n$
and is invariant under conjugation by any element $\s\in S_n$:
$\a \mapsto \s^{-1} \a \s$.  The number of traces of length $k$ in an
operator is just the number of cycles of length $k$ in $\a$.  In this
way the different operators with $n$ fields map to the different
conjugacy classes of $S_n$, which are in one-to-one correspondence
with the partitions of $n$ into integer parts.\footnote{We shall be
  loose in our notation: $[n_1,n_2, \cdots n_k]$ refers both to the
  ordered partition of $n$ that defines a conjugacy class of $S_n$ and
  to the canonical permutation within that conjugacy class $(12 \cdots
  n_1)(n_1+1 \cdots n_1+n_2) \cdots (\cdots n)$.}

The scalar propagator
\begin{equation}
  \corr{(X^{\dagger})^i_j (x) \; X^k_l(y)} = \delta^i_l \delta^k_j \frac{1}{(x-y)^2} \label{eq:basicXcor}
\end{equation}
can now be used to compute correlation functions of these operators.
We choose the positions and coordinate frames of the operators in the
two-point function to remove the spacetime dependence so that we can
focus on the combinatorics of the contraction of the gauge
indices.\footnote{This choice is akin to the 2d Zamolodchikov metric:
  we put the operators at opposite poles of $S^4$ in coordinate frames centred about each pole respectively.}  The simplest
example is the two-point function of two single-trace
operators\footnote{Higher genus corrections are computed in Appendix
  Section \ref{sec:highergenus}.}
\begin{equation}
  \corr{ \tr(X^{\dagger n}) \tr(X^n)} = nN^n\left\{1 + \left[ \binom{n}{3} +\binom{n}{4} \right]\frac{1}{N^{2}} + \cO\left(\frac{1}{N^4}\right)  \right\} \label{eq:FullGT}
\end{equation}

We will now review how, once we map the operators to conjugacy classes
of $S_n$, all the $\frac{1}{N}$ corrections are captured by class
functions of $S_n$ acting on these states.  All the interactions can
be encoded in the element $\O_n$ familiar from the $\frac{1}{N}$ expansion of 2d Yang-Mills \cite{2dYM,9411210}
\begin{align}
  \O_n & = \sum_{\s \in S_n} N^{C(\s)-n} \;\;\s \nn \\
& = 1 + \frac{1}{N} \S_{[2]} + \frac{1}{N^2} \left(  \S_{[3]}+\S_{[2,2]} \right) + \frac{1}{N^3} \left(  \S_{[4]}+ \S_{[3,2]}+\S_{[2,2,2]} \right) + \cO\left(\frac{1}{N^4}\right)  
\end{align}
$C(\s)$ is the number of cycles in $\s$.  The sum over transpositions
$[2,1^{n-2}] = [2,\overbrace{1,1,\cdots 1}^{n-2}]$ (a total of $n-1$
cycles) is written $\S_{[2]}$ and similarly $\S_{[3,2]}$ sums over all
permutations of the form $[3,2,1^{n-5}]$ (a total of $n-3$ cycles).
Since each $\S_C$ is of the form $\S_{C} \propto \sum_{\rho \in S_n} \rho\, \s
\,\rho^{-1}$ for some $\s$ in the conjugacy class $C$, $\S_C$ commutes
with every element of $S_n$ and is hence in the centre of $S_n$.

The non-planar two-point function of two single-trace operators can
now be written
\begin{equation}
  \corr{ \tr(X^{\dagger n}) \tr(X^n)} = N^n\;\bra{n} \O_n \ket{n}
\end{equation}
Furthermore for large $n$ (but still $n < N^{\frac{1}{2}}$ so that
mixing between operators with different trace structures is
suppressed)\footnote{The limit of $n$ large with $\frac{n^2}{N} < 1$
  fixed is the same as the BMN limit \cite{0202021} for the half-BPS
  sector.  However, as we will show later, the exponentiation of
  $\O_n$ works in any sector, even when the number of fields $n$ is
  not related to the $R$-charge $J$ as it is for the half-BPS sector
  where $n=\Delta=J$.\label{foot:BMN}} the dominant part of $\O_n$
exponentiates \cite{0209215}
\begin{equation}
  \O_n \to \exp \left( \frac{1}{N} \S_{[2]} \right)
\end{equation}
Geometrically this corresponds to the factorisation of higher genus
two-dimensional surfaces with marked points into three-punctured
spheres.

\subsection{The Hilbert space and its inner product}

The half-BPS Hilbert space is defined by conjugacy classes of $S_n$
\begin{equation}
 \ket{n_1,n_2, \cdots n_k} \equiv \tr([n_1,n_2, \cdots n_k]\; X) = \tr(X^{n_1})\tr(X^{n_2})\cdots \tr(X^{n_k})
\end{equation}
Its conjugate is
\begin{equation}
 \bra{n_1,n_2, \cdots n_k} \equiv \tr([n_1,n_2, \cdots n_k]\; X^\dagger) = \tr(X^{\dagger n_1})\tr(X^{\dagger n_2})\cdots \tr(X^{\dagger n_k})
\end{equation}
The structure of the Hilbert space is exactly the same as that in
\cite{0209215}.

Define an inner product by taking the planar part of the two-point function
\begin{align}
    \braket{m_1, \dots m_l}{n_1, \dots n_k} & = \frac{1}{N^n}\corr{\tr(X^{\dagger m_1})\cdots \tr(X^{\dagger m_l}) \;\;\tr(X^{n_1})\cdots \tr(X^{n_k})}_{\textrm{planar}} \nn \\
&= | \Sym([n_1,\dots n_k])| \;\; \delta_{[m_1, \dots m_l] = [n_1, \dots n_k]} \label{eq:firstinner}
\end{align}
The conjugacy class $[\vec{m}]$ must be the same as $[\vec{n}]$
for the inner product to be non-zero.  This means that the trace
structures must be the same.  When it is non-zero the value is the
size of the symmetry group of $[\vec{n}]$, i.e. the group of
permutations that leave $[\vec{n}]$ invariant under conjugation.  If
$[\vec{n}]$ has $i_k$ $k$-cycles then its size is
$i_1!1^{i_1}i_2!2^{i_2} \cdots i_n!n^{i_n}$.  The factors $k^{i_k}$
corresponds to the cyclic symmetry $\Z_k$ of each cycle while $i_k!$
is the permutation factor for the $i_k$ identical $k$-cycles.  

By taking the planar part of the two-point function we have in mind
the 2d Zamolodchikov metric where we get an inner product for a 2d
theory by taking the two-point function on the sphere $S^2$ with
operators at opposite poles.

In the simplest example for single traces
\begin{equation}
  \braket{n}{n} =\frac{1}{N^n}\corr{ \tr(X^{\dagger n}) \tr(X^n)}_{\textrm{planar}} = n
\end{equation}
where $n$ is the size of the cyclic group $\Z_n$ that leaves the
$n$-cycle $(12\cdots n)$ invariant.

For two-particle states
\begin{equation}
  \braket{n_1,n_2}{n_1,n_2} = \left\{\begin{array}{l} n_1n_2\quad \textrm{ for } \quad n_1 \neq n_2 \\
2n_1n_2\quad \textrm{ if } \quad n_1 = n_2  \end{array}  \right.
\end{equation}
which is just the product of two single-trace inner products.  For
$n_1=n_2$ we must be more careful: there is an extra factor of $2$
because the symmetry group now includes the exchange of the two
identical cycles.

\subsection{Planar three-point function}\label{sec:3ptfun}

The leading planar contribution to the extremal three-point function
\cite{9908160} of single-trace half-BPS operators is the same as that
of the two-point function between a single-trace operator and a
two-trace operator\footnote{The spacetime dependence of the
  three-point function is generically $(x-y)^{-2n_1}(x-z)^{-2n_2}$.
  However, because there is no interaction between $\tr(X^{n_1})$ and
  $\tr(X^{n_2})$ we can take these operators to be close together and
  treat the three-point function like the two-point function between
  $[\tr(X^{\dagger n})](x)$ and
  $[\tr(X^{n_1})\tr(X^{n_2})](y)$.\label{ft:caution}}
\begin{equation}
  \corr{ \tr(X^{\dagger n}) \tr(X^{n_1})\tr(X^{n_2})} = nn_1n_2N^{n-1} + \cO(N^{n-3})
\end{equation}
for $n_1 + n_2=n$. The inner product between these states vanishes
$\braket{n}{n_1,n_2}=0$ so we must introduce an interaction vertex of
order $\frac{1}{N}$ to get this non-trivial result.  Define an element
of the algebra of the symmetric group $\Q S_n$ (elements of $S_n$ with
rational coefficients) by a sum over all of the $\binom{n}{2}$
two-cycles, i.e. all elements of $S_n$ in the conjugacy class
$[2,1^{n-2}]$
\begin{equation}
  \S_{[2]} = \sum_{i<j} (ij) = \sum_{\s \in [2,1^{n-2}]} \s
\end{equation}
It acts on $\ket{n_1,\dots n_k}$ by left action on the permutation defining the state
\begin{equation}
  \S_{[2]} \ket{n_1,\dots n_k} = \sum_{\s \in [2,1^{n-2}]} \;\;\;\ket{\s[n_1,\dots n_k]}
\end{equation}
Applying it to $\ket{n}$, $\S_{[2]}$ splits the single trace into all possible double traces\footnote{In more detail: consider the
  transposition $(ij)$ acting on the canonical $n$-cycle $(12\cdots i
  \cdots j \cdots n)$.  Just cycling around the elements, the
  $n$-cycle is identical to $(i\cdots j \cdots n12\cdots)$.  Denote
  the first sequence abbreviated to $\cdots$ by $s_1 = i+1\cdots j-1$
  and the second by $s_2 = j+1\cdots n12\cdots i-1$.  The computation
  is now $(ij)(i\,s_1\,j\,s_2) = (i\,s_1)(j\,s_2)$.\label{eq:ijdetail}}
\begin{equation}
  \S_{[2]}\ket{n} = \sum_{k=1}^{n-1} \frac{n}{2} \ket{ n-k,k}
\end{equation}
If $n_1 \neq n_2$ we get the resulting vertex for the three-point function
\begin{equation}
  \bra{n_1,n_2}  \S_{[2]}\ket{n} = \frac{n}{2} \braket{n_1,n_2}{n_1,n_2}+ \frac{n}{2} \braket{n_1,n_2}{n_2,n_1} = nn_1n_2
\end{equation}
For  $n_1 = n_2$ we arrive at the same result
\begin{equation}
  \bra{n_1,n_1}  \S_{[2]}\ket{n} = \frac{n}{2} \braket{n_1,n_1}{n_1,n_1} = nn_1n_2
\end{equation}
To check the symmetry under conjugation, $\S_{[2]}$ also joins traces when
acting on the double trace
\begin{equation}
  \S_{[2]} \ket{n_1,n_2} = n_1n_2 \ket{n} + \sum_{k=1}^{n_1-1} \frac{n_1}{2} \ket{ n_1-k,k,n_2} + \sum_{k=1}^{n_2-1} \frac{n_2}{2} \ket{n_1,n_2-k,k} \label{eq:S2onn1n2}
\end{equation}
so that the result is preserved
\begin{equation}
  \bra{n}\S_{[2]} \ket{n_1,n_2} = nn_1n_2
\end{equation}

Including the $N$ dependence the 3-point function vertex is
\begin{equation}
  \frac{1}{N} \S_{[2]} \label{eq:3ptvertex}
\end{equation}

\subsection{Torus two-point function}

We now use the same formalism to compute the first non-planar
correction to the single-trace two-point function \eqref{eq:FullGT}.
The lowest genus Riemann surface on which the fat graphs of these
Feynman diagrams can be drawn is the torus.

For a correction of $\frac{1}{N^2}$ the cut-and-join operators come
from a sum over the elements of $S_n$ with $n-2$ cycles.  These
elements are composed of two transpositions, i.e. either two 2-cycles
or one 3-cycle. If we define $\S_{[3]}$ to be the sum of all
permutations with one 3-cycle (a sum of $2\binom{n}{3}$ terms) and
$\S_{[2,2]}$ to be the sum of all permutations with two 2-cycles (a
sum of $3\binom{n}{4}$ terms) then we find
\begin{align}
  \S_{[3]} \ket{n} &= \binom{n}{3} \ket{n} + \cdots\nn \\
  \S_{[2,2]} \ket{n} &= \binom{n}{4} \ket{n} + \cdots  \label{eq:2cuts}
\end{align}
Roughly, each of these is akin to two actions of $\S_{[2]}$ so that in
the first term a single trace has split and rejoined.  In the $\cdots$
are three-particle states from two splittings (see Appendix Section
\ref{sec:4point} for these terms and their use in the extremal
four-point function).

The torus two-point function is then given by
\begin{align}
  \bra{n} \left( \S_{[3]}+\S_{[2,2]} \right) \ket{n} & = n  \left[ \binom{n}{3} +\binom{n}{4} \right]  \label{eq:correct}
\end{align}
This agrees with the calculation in
\eqref{eq:FullGT}.  Including the $N$ dependence the correct vertex is
\begin{equation}
  \frac{1}{N^2} \left( \S_{[3]}+\S_{[2,2]} \right) \label{eq:2vertex}
\end{equation}

\subsection{Relation to worldsheet models}\label{sec:worldsheet}

\begin{figure}[t]
\begin{center}
\resizebox{!}{5cm}{\includegraphics[trim=0 0 0 0 ]{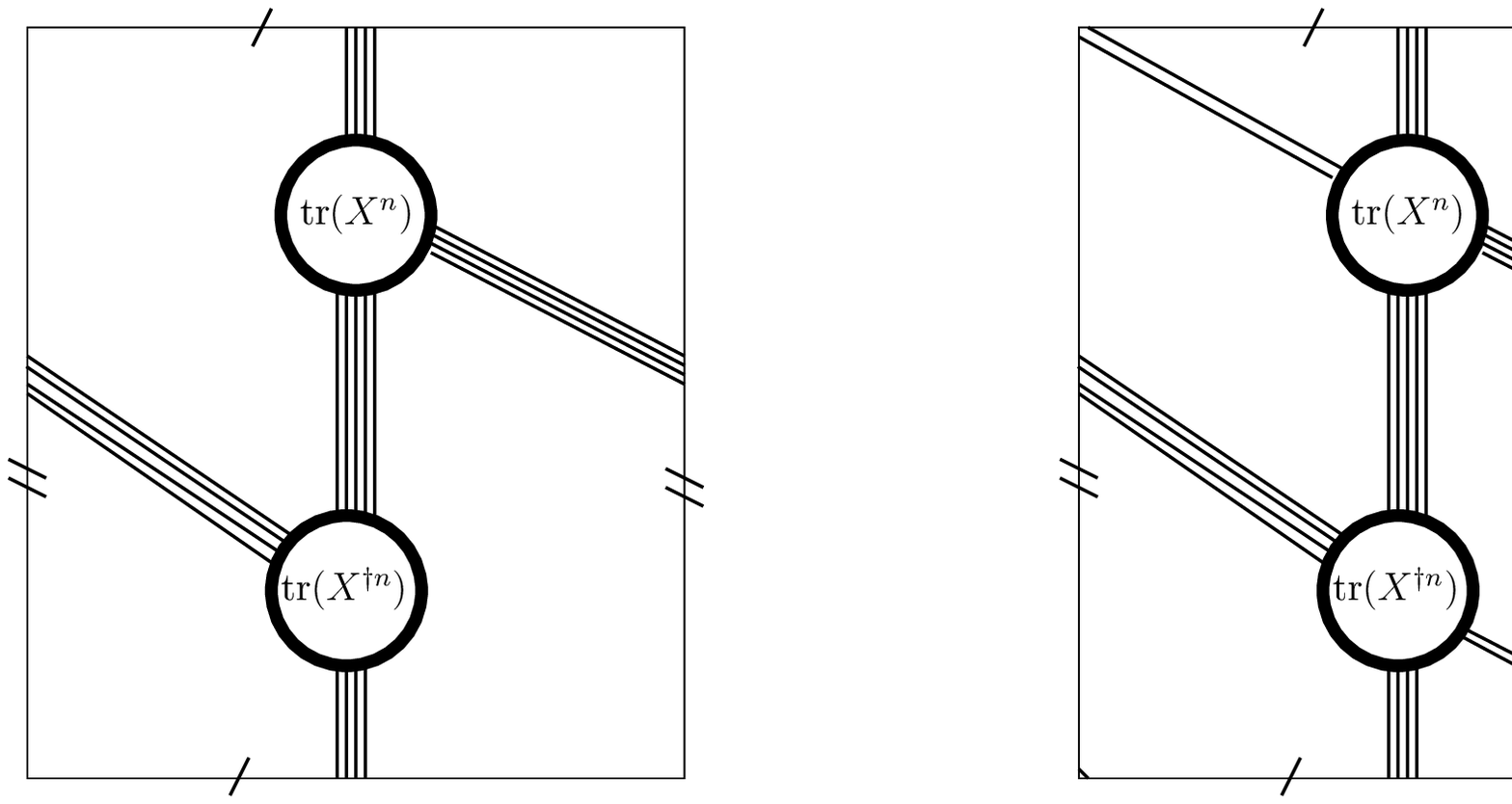}}
\caption[]{The two different bunchings of propagators that can be
  drawn on the torus with no crossing: three bunchings from $\S_{[3]}$
  on the left and four bunchings from $\S_{[2,2]}$ on the right.  This
  figure is copied from Figure 1 of \cite{0205033} and Figure 2 of
  \cite{0205089}.}\label{fig:torworld}
\end{center}
\end{figure}

What do the combinatorial numbers in \eqref{eq:correct} capture in the
dual worldsheet theory?  For two ordered circular chains of fields
(the single traces in index space) the binomials count the number of
ways of Wick contracting the fields on the two-dimensional torus so
that none of the propagators cross \cite{0205033,0205089} (excluding
those diagrams which can also be drawn on the sphere).  Where
propagators run parallel to each other they can be bunched together
into homotopic groups.  On the torus this can be done into either
three groups, of which there are $\binom{n}{3}$ ways of bunching them,
or four groups, of which there are $\binom{n}{4}$ ways. See Figure
\ref{fig:torworld} copied from \cite{0205033,0205089}.

The bunching of propagators into homotopic groups to form `skeleton
graphs' is an important feature of Gopakumar's mapping of the free
theory Feynman diagrams to worldsheet correlators
\cite{gopakumar}.  The vertex \eqref{eq:2vertex}
is also the relevant one for the extremal four-point function.  The
two classes of bunchings for the four-point function, called the $Y$
and lollipop diagrams in \cite{freehigher}, also
correspond to $\S_{[3]}$ and $\S_{[2,2]}$ respectively, see Appendix
Section \ref{sec:4point}.  We are therefore lead to make the
conjecture:
\begin{equation}
  \boxed{\parbox{12cm}{For each extremal $k$-point function on a genus
      $g$ surface the different homotopic bunchings of propagators
      correspond to the different conjugacy classes that appear at
      order $\frac{1}{N^{k+2g-2}}$ in $\O_n$.}} \label{eq:gopconj}
\end{equation}
The number of bunches for $\S_C$ is given by the sum of the
non-trivial parts of $C$. For example, $\S_{[2,2]}$ splits
$\ket{n}$ into $2+2=4$ bunches because the typical element of $C$, say
the permutation $(ij)(kl)$, reorders the $n$-cycle $(12\cdots n)$ into
four separate pieces across one $n$-cycle or three smaller cycles.

Berkovits' model for the free field theory \cite{0806.1960} also captures
the non-crossing of propagators.

\subsection{Factorisation of the torus two-point function into planar three-point functions}

\begin{figure}[t]
\begin{center}
\resizebox{!}{2.4cm}{\includegraphics[trim=10 0 -85 0 ]{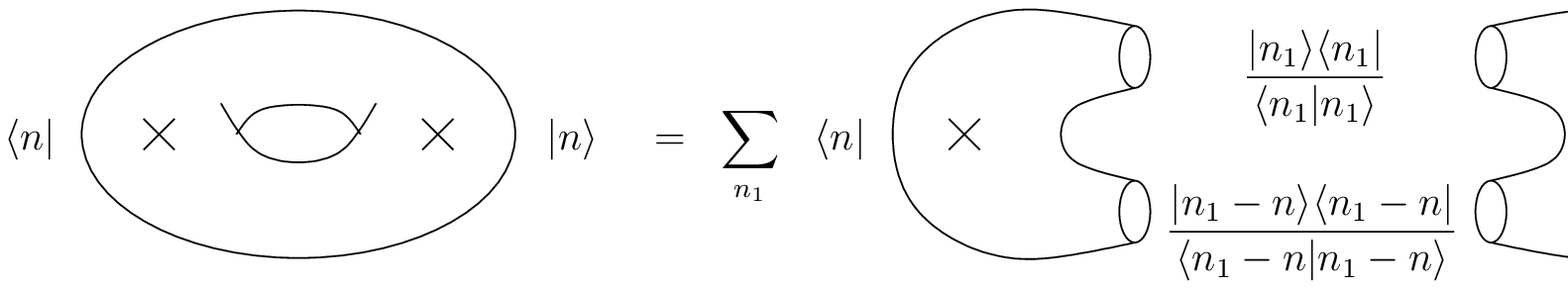}}
\caption[]{Torus factorisation.}\label{fig:torus}
\end{center}
\end{figure}

An interesting feature of the torus two-point function is that it
factorises into the product of two planar three-point functions in the
limit when $n$ is large (so that the operators have many fields).
This factorisation was first pointed out in investigations
\cite{0205089,0206248} of the BMN limit \cite{0202021}\textsuperscript{\ref{foot:BMN}}, in particular using the string bit
formalism \cite{0209215} where these results appeared in this
language. Such a factorisation takes the form of a sum over two sets
of intermediate states (see Figure \ref{fig:torus})
\begin{equation}
 \bra{n}\S_{[2]} \S_{[2]} \ket{n} =  \sum_{n_1 \leq \frac{n}{2}}  \bra{n} \S_{[2]} \frac{\ket{n_1,n-n_1}\bra{n_1,n-n_1}}{\braket{n_1,n-n_1}{n_1,n-n_1}} \S_{[2]} \ket{n}  \label{eq:factorise}
\end{equation}
Factorisations like this are familiar from two-dimensional topological
field theories, but this association may be na\"ive given that the
factorisation only takes place in index space: we have ignored the
non-trivial spacetime dependencies of the correlation functions,
particular the three-point function.\textsuperscript{\ref{ft:caution}}

To see the equality in \eqref{eq:factorise} note that only
two-particle states appear in the image of $\S_{[2]}$ acting on
$\ket{n}$ so that the projector
\begin{equation}
  \sum_{n_1 \leq \frac{n}{2}}
\frac{\ket{n_1,n-n_1}\bra{n_1,n-n_1}}{\braket{n_1,n-n_1}{n_1,n-n_1}}
\end{equation}
has a trivial action on $\S_{[2]}\ket{n}$. Thus we can just insert it
in the middle of $\bra{n} \S_{[2]}\S_{[2]}\ket{n}$ to get
\eqref{eq:factorise}.  The half-BPS operators only mix with other
half-BPS operators in the extremal three-point function, which is why
the sum over intermediate states truncates to the half-BPS sector.

Next we want to show that \eqref{eq:factorise} agrees with the correct
result \eqref{eq:correct} in the large $n$ limit.
\begin{equation}
  \S_{[2]}\S_{[2]} = \binom{n}{2}  + 3\S_{[3]}+ 2 \S_{[2,2]} \label{eq:goat}
\end{equation}
so that if we use the results for the actions on $\ket{n}$ in
\eqref{eq:2cuts} we find
\begin{align}
 \bra{n} \S_{[2]}\S_{[2]}\ket{n} & =n \left[ \binom{n}{2} + 3\binom{n}{3} + 2 \binom{n}{4}\right] \label{eq:gt}
\end{align}
This is not exactly the same as the correct field theory result
\eqref{eq:correct}.  However in the large $n=\Delta = J$ limit the
first terms in both \eqref{eq:correct} and \eqref{eq:gt} are
suppressed and the leading behaviour $\binom{n}{4} \sim
\frac{n^4}{4!}$ is the same for both up to a factor of 2.  Essentially
the term $\S_{[2,2]}$ dominates over the other terms because its size
is greatest, of order $\binom{n}{4}$.  This result generalises as we
see in the next section.

\subsection{Factorisation for the full string cut-and-join operator}\label{sec:factor}

The full string cut-and-join operator is \cite{0209215}
\begin{align}
  \O_n & = \sum_{\s \in S_n} \frac{1}{N^{n-C(\s)}} \;\;\s \nn \\
  & = 1 + \frac{1}{N} \S_{[2]} + \frac{1}{N^2} \left( \S_{[2,2]}+ \S_{[3]} \right) + \frac{1}{N^3} \left( \S_{[2,2,2]}+ \S_{[3,2]}+ \S_{[4]} \right) + \cO\left(\frac{1}{N^4}\right)
\end{align}
 $C(\s)$ is the number of cycles in $\s$.  The negative power of $N$
is also the minimum number $T(\s)$ of transpositions it takes to build each
element $\s$, which is related to $C(\s)$ by
\begin{equation}
  T(\s) = n-C(\s)
\end{equation}

A check on $\O$ is that it reproduces the full non-planar two-point
function. For operators built with two permutations $\a,\a' \in S_n$
the two-point function is a sum over all permutations $\tau\in S_n$ of
Wick contractions of the fields that gives \cite{0111222}
\begin{align}
  \corr{ \tr(\a'\, X^{\dagger n}) \tr(\a\, X^n)} & = \sum_{\tau \in S_n} N^{C(\a\tau\a'\tau^{-1})}\nn \\
& =  \sum_{\s \in S_n}\sum_{\tau \in S_n} N^{C(\s)} \delta(  \s \a\tau\a'\tau^{-1} = 1) \label{eq:correquib}
\end{align}
To see why the second equality holds, consider the coefficient of a
particular power $h$ of $N$.  To contribute to this coefficient
$\a\tau\a'\tau^{-1}$ must have $h$ cycles.  To test this condition,
multiply $\a\tau\a'\tau^{-1}$ by all possible permutations $\s$ with
$h$ cycles and see if $\s\a\tau\a'\tau^{-1} = 1$.  This is
equivalent to the condition that $\tau\a'\tau^{-1}$ is the inverse of
$\s\a$, which because we sum over $\tau \in S_n$ happens
$|\Sym(\a)|$ times (if $\s \a$ is in the same conjugacy class as
$\a'$).  But this is exactly what we do when we insert $\O$ in the
inner product: for $N^h$ we multiply $\ket{\a}$ by all possible
permutations with $h$ cycles then $\bra{\a'}\O\ket{\a}$ is only
non-zero if $\s\a$ is in the same conjugacy class as $\a'$ (and
the inner product gives us a factor of $|\Sym(\a)|$).  Thus we get
\begin{equation}
  \corr{ \tr(\a'\, X^{\dagger n}) \tr(\a\, X^n)} = N^n\bra{\a'}\O_n\ket{\a}
\end{equation}

Taking the large $n$ limit at each order in $\frac{1}{N}$ we find that $\O$ exponentiates
\begin{equation}
  \boxed{\O_n \to \sum_{i} \frac{1}{i!}\frac{1}{N^i}\left(  \S_{[2]} \right)^i = \exp\left(\frac{1}{N} \S_{[2]} \right)}\label{eq:expvertex}
\end{equation}
This limit for the operator is meant in the sense that every matrix
element $\bra{\vec{m}}\O\ket{\vec{n}}$ satisfies this limit. The
number of elements in $\S_{[2]}$ is $\binom{n}{2}\sim n^2$
so that for $\O_n$ to remain non-vanishing requires the BMN
ratio\textsuperscript{\ref{foot:BMN}} $g_2 \equiv \frac{n^2}{N}<1$ to be
fixed in the large $n,N$ limit.  The exponentiation of $\O$ can in
fact be made exact for any $n$ by adding extra terms, see equation
\eqref{eq:exactexp} in Appendix \ref{sec:on}, from which it is 
clear that the corrections to \eqref{eq:expvertex} are subleading in
$g_2$.

Equation \eqref{eq:expvertex} holds because the size of the conjugacy
class $\S_{[2^i]}$ grows like $\binom{n}{2i} \sim n^{2i}$ which is a
factor of $n$ larger than any other operator that appears at order
$N^{-i}$ (see Appendix Section \ref{sec:highergenus} on the general
form of higher genus operators).  Similarly it dominates $ \left(
\S_{[2]} \right)^i$ in the large $n$ limit
\begin{equation}
  \left(  \S_{[2]} \right)^i \to i!\; \S_{[2^i]}
\end{equation}
This equivalent behaviour gives equation \eqref{eq:expvertex}.

The exponentiated term in equation \eqref{eq:expvertex} is the
three-point function vertex studied in Section \ref{sec:3ptfun}.  Thus
the large $n$ limit exponentiation corresponds geometrically to the
factorisation of higher genus correlation functions into the
three-point functions which map to three-puncture spheres
\cite{0209215}.  For small $n$ the exponentiation of $\O$ still holds,
see equation \eqref{eq:exactexp} in Appendix \ref{sec:on}, but there
are higher order terms so we lose the geometrical interpretation.

\subsection{Relation to Schur polynomials and $U(N)$ Casimirs}\label{sec:schurpoly}

At finite $N$ linear relations appear between trace operators with
large dimension $\D \geq N$.  This is known as the Stringy Exclusion
Priniciple.  A well-defined operator can be constructed by `Fourier
transforming' to representations $R$ of $S_n$.  Because it is used to
trace the gauge indices, the Young diagram $R$ also labels a
representation of $U(N)$.  This limits the number of rows to $N$,
which neatly encodes the Stringy Exclusion Principle.  Columns and
rows correpond to giant gravitons in the bulk \cite{0111222} and more
general geometries via the LLM description \cite{0409174}.

The Schur polynomial is a linear combination of the trace operators where the coefficient is the $S_n$ character $\chi_R(\s)$ of $R$
\begin{equation}
  \ket{R} = \frac{1}{n!} \sum_{\s \in S_n}\chi_R(\s) \ket{\s}
\end{equation}
This basis is not only diagonal in the planar inner product
\begin{equation}
  \braket{S}{R} = \delta_{RS}
\end{equation}
but also in the full non-planar two-point function.  To see this
compute the action of a generic cut-and-join operator $\S_C$ where $C$
labels a conjugacy class of $S_n$
\begin{equation}
\S_{C} \ket{R}  = \frac{\chi_R(\S_{C})}{d_R} \ket{R}
\end{equation}
It acts simply by multiplication. This is a result of Schur's lemma,
because the $\S_{C}$ are in the centre of the group $S_n$.  $d_R$ is
the symmetric group dimension.  This means the non-planar two-point function is
\begin{align}
  \bra{S} \O_n \ket{R}  & =  \frac{\chi_R(\O_n)}{d_R} \braket{S}{R} \nn \\
  & =  \frac{n! \dim R}{N^n d_R} \;\;\delta_{RS}
\end{align}
We have used formula \eqref{eq:Ondim} for the $U(N)$ dimension $\dim R$.

The commuting $\S_{C}$ are linear combinations of the commuting higher
Hamiltonians defined in \cite{0111222,0807.3696} which correspond to
Casimirs of the gauge group $U(N)$.  They distinguish the different
Schur polynomials $\ket{R}$ by their eigenvalues.  For example the
second and third Casimirs in the representation $R$ are
\begin{align}
  C_2(R) & = nN +2\frac{ \chi_R(\S_{[2]})}{d_R} \nn \\
  C_3(R) & = n(n-1)+ nN^2 +4N\frac{ \chi_R(\S_{[2]})}{d_R} +3\frac{ \chi_R(\S_{[3]})}{d_R} 
\end{align}
See Section 4.10.1 of the review of two-dimensional
Yang-Mills \cite{9411210} for more examples.

These charges have been identified in the dual gravitational
description with asymptotic multipole moments of the spacetime
\cite{0602263}.

\section{Extension to general representations in the free theory}\label{sec:general}

In the previous section we used cut-and-join operators to compute
non-planar correlation functions of half-BPS operators.  These
operators were constructed from a single complex scalar $X$.  In this
section we allow more than one field from the theory.  Non-trivial
symmetrisations are now possible, such as the commutator of two
complex fields $[X,Y]$.  For non-planar calculations in the free
theory only the symmetrisation of the operator is important, which is
indexed by a Young diagram of $n$ boxes (for $n$ fields).  This is
equivalent to treating the global symmetry group as $U(\infty)$ where
the fundamental multiplet is just the infinite number of single fields
in the singleton representation of $PSU(2,2|4)$, including
derivatives.  These representations are the same as those for the
higher spin theory, which are organised into YT-pletons in
\cite{0405057}.

Only at one-loop do the specific $PSU(2,2|4)$ representations become
important.  Each $U(\infty)$ representation breaks down into an
infinite number of $PSU(2,2|4)$ representations. In \cite{0806.1911}
it was shown how to do this using Schur-Weyl duality for $SO(6)$ and
$SO(2,4)$.  The general story is outlined in Section
\ref{sec:gengroup}.

To start with we focus on the $U(2)\subset SU(4)_R \subset PSU(2,2|4)$
example but the language will be kept general enough that the
extension to $U(\infty)$ is obvious.  Some of the initial parts of
this analysis were also carried out in \cite{0404066,0604174}.

\subsection{Not organising into representations of $U(2)$}

In the $U(2)$ subsector of the global symmetry group there are two
species of field $X$ and $Y$.  To compute the two-point function it
is necessary to modify the inner product appropriately, but the
non-planar corrections are still given by $\O_n$.  We borrow heavily
from \cite{0711.0176} where the non-planar correlation functions for
the free theory were calculated.

To begin with we won't organise the operators into representations of
$U(2)$.  The state space is defined by a field content of $\m_1$ $X$'s
and $\m_2$ $Y$'s and a permutation $\a\in S_n$
\begin{equation}
  \ket{\vec{\m},\a} \equiv \tr(\a\, X^{\m_1} Y^{\m_2}) = X^{i_1}_{i_{\a(1)}} \cdots  X^{i_{\m_1}}_{i_{\a(\m_1)}} Y^{i_{\m_1+1}}_{i_{\a(\m_1+1)}}\cdots Y^{i_{\m_1+\m_2}}_{i_{\a(\m_1+\m_2)}} \label{eq:basiop}
\end{equation}
The states are not defined by conjugacy classes of $\a$, but up to a
new equivalence class $[\a]$ defined by the relation
\begin{equation}
  \a \sim \rho^{-1} \a \rho \quad \quad \rho \in S_{\m_1}\times S_{\m_2} \label{eq:alphsymmcong}
\end{equation}
Under this conjugation the operator $\tr(\a\, X^{\m_1} Y^{\m_2})$
remains invariant because the canonical choice of ordering of
$X^{\otimes \m_1}\otimes Y^{\otimes \m_2}$ is unchanged by $\rho \,
X^{\otimes \m_1}\otimes Y^{\otimes \m_2}\, \rho^{-1}$ for $\rho \in
S_{\m_1}\times S_{\m_2}$ (see equation \eqref{eq:snaction} for an example of this kind of $S_n$ action).\footnote{We can cycle permutations around
  the trace so that $\tr(\a \rho \, X^{\otimes \m_1}\otimes Y^{\otimes
    \m_2}\, \rho^{-1}) = \tr( \rho^{-1}\a \rho \, X^{\otimes
    \m_1}\otimes Y^{\otimes \m_2})$.}

In the bra the order of the fields is reversed so that it is defined
with the inverse $\a^{-1}$
\begin{equation}
  \bra{\vec{\m},\a} \equiv \tr(\a^{-1}\, X^{\dagger \m_1} Y^{\dagger\m_2})
\end{equation}
The symmetry \eqref{eq:alphsymmcong} is reflected in the planar inner
product, where the non-zero value is given by the size of the
intersection of $S_{\m_1}\times S_{\m_2}$ with $\a$'s symmetry group
\begin{equation}
  \braket{\vec{\m}\,{}',\a'}{\vec{\m},\a} = | \Sym(\a) \cap ( S_{\m_1}\times S_{\m_2}) | \;\;\delta_{\vec{\m}\vec{\m}\,{}'} \delta_{\a\sim\a'}\label{eq:notorginn}
\end{equation}
The cut-and-join operators act by left-multiplication on $\a$, just as before
\begin{equation}
  \S_{[2]} \ket{\vec{\m},\a}  = \sum_{\s \in [2,1^{n-2}]} \;\;\;\ket{\vec{\m},\s\a}
\end{equation}
For example take the single-trace state
\begin{equation}
  \ket{\vec{\m}=(2,2),\a=(1324)} = X^{i_1}_{i_3}X^{i_2}_{i_4}Y^{i_3}_{i_2}Y^{i_4}_{i_1} = \tr(XYXY)
\end{equation}
The action of $\S_{[2]}$ splits this state into three different kinds of double-trace operator
\begin{align}
  \S_{[2]}\ket{\vec{\m},\a=(1324)} & = 2\ket{\vec{\m},\a=(13)(24)} + 2\ket{\vec{\m},\a=(1)(324)}+2\ket{\vec{\m},\a=(132)(4)} \nn \\
& = 2\tr(XY)\tr(XY) + 2\tr(X) \tr(YXY) + 2 \tr(XYX)\tr(Y)
\end{align}
Once again $\frac{1}{N}\S_{[2]}$ defines the planar three-point
function vertex.\footnote{Note that the only double-trace operator
  with $\vec{\m}=(2,2)$ which doesn't appear here, $\tr(XX)\tr(YY)$,
  has no overlap with $\tr(XYXY)$ until $N^{-3}$.}

Using the inner product \eqref{eq:notorginn} the non-planar expansion
of the correlation function is given by $\O_n$ by the same line of
reasoning below \eqref{eq:correquib}
\begin{equation}
  \corr{ \tr(\a^{\prime -1}\, X^{\dagger \m_1} Y^{\dagger \m_2}) \tr(\a\,X^{\m_1} Y^{\m_2} )} = \sum_{\s \in S_{\m_1} \times S_{\m_2}} N^{C(\a\s\a^{\prime -1}\s^{-1})}  =  N^n\,\bra{\vec{\m},\a'}\, \O_n\, \ket{\vec{\m},\a} 
\end{equation}
The exponentiation of the S-matrix $\O_n = \exp \left( \tfrac{1}{N}
\S_{[2]}\right)$ and hence the factorisation into planar three-point
functions for large $n= \m_1 + \m_2$ follows exactly the same
reasoning as in the half-BPS sector in Section \ref{sec:factor}.

\subsection{Organising into representations}\label{sec:organrep}

In this section we show how to organise operators built from $X$s and
$Y$s with a given trace structure $\a$ into representations $\L$ of
$U(2)$.  $\L$ is a Young diagram with $n$ boxes and at most two rows,
which we write $\L \in P(n,2)$ to indicate that $\L$ is a partition of
$n$ into at most 2 parts. $\L$ is also a representation of $S_n$ which
shows how the operator is symmetrised.  The cut-and-join technology
follows through because these representations are just linear
combinations of the operators constructed above.

To organise the fields into representations of $U(2)$ first strip the
fields of their gauge indices so that they live in an abstract tensor
space of $U(2)$.  If $V_{\bf 2}$ is the fundamental of $U(2)$ consisting of
the two complex scalars $X$ and $Y$ then $n$ copies of this rep
$V_{\bf 2}^{\otimes n}$ can be sorted into irreps of $U(2) \times S_n$
using Schur-Weyl duality
\begin{equation}
  V_{\bf 2}^{\otimes n} = \sum_{\L \in P(n,2)} V_{\L}^{U(2)} \otimes V_{\L}^{S_n} \label{eq:SWdecomp}
\end{equation}
In each summand the Young diagram $\L$ for the $U(2)$ representation
and the $S_n$ representation is the same.  This diagonal decomposition
works because $U(2)$ and $S_n$ have a commuting action on $V_{\bf
  2}^{\otimes n}$.  To implement the decomposition concretely we
define an action of $\s \in S_n$ on $V_{\bf 2}^{\otimes n}$ by
conjugation.  For example on an element of $V_{\bf 2}^{\otimes 4}$ the
permutation $\s = (13) \in S_4$ has the action
\begin{equation}
  \s\;\; X\otimes X \otimes Y\otimes Y\;\;\s^{-1} = Y \otimes X \otimes X \otimes Y \label{eq:snaction}
\end{equation}
The decomposition \eqref{eq:SWdecomp} means that a complete basis is given by the linear combination of elements of $V_{\bf 2}^{\otimes n}$ (to avoid confusion we will keep the standard braket notation $\ket{\cdots}$ for gauge-invariant states; for other states  which live in general tensor spaces we will use $\mket{\cdots}$)
\begin{equation}
  \mket{\L,M,a} \equiv\frac{1}{n!} \sum_{\s \in S_n}\; B_{ b\b}^{\vec{\m}}\; D^{\L}_{ab}(\s) \;\;\;\;\; \s \;\;\overbrace{X\otimes X \otimes \cdots X}^{\m_1}\otimes \overbrace{Y\otimes Y \otimes \cdots Y}^{\m_2}\;\;\s^{-1} \label{SWstate}
\end{equation}
The operator is labelled by
\begin{itemize}
\item A Young diagram $\L$ with $n$ boxes and at most 2 rows;
\item The label $M = \{\vec{\m},\b\}$ which tells us which $U(2)$
  state from the representation $V_{\L}^{U(2)}$ it is.  $\vec{\m}$
  labels the field content and $\b$ the semi-standard tableau for
  $\vec{\m}$ and $\L$;
\item $a$ is the $S_n$ state for $V_{\L}^{S_n}$.
\end{itemize}
$D^{\L}_{ab}(\s)$ is the orthogonal matrix for $\s \in S_n$ in the
representation $\L$; in \eqref{SWstate} there is Einstein summation
over the $V_{\L}^{S_n}$ state $b$.  The parameter $B_{b\b}^{\vec{\m}}$
is a branching coefficient for $S_n \to S_{\m_1}\times S_{\m_2}$
described in more detail below.

Reintroducing the gauge indices and tracing with $\a$ gives a
gauge-invariant operator
\begin{equation}
  \ket{\L,M,a;\a} \equiv \frac{1}{n!} \sum_{\s \in S_n}\; B_{ b\b}^{\vec{\m}} \; D^{\L}_{ab}(\s)\;\;  \tr(\a\s X^{\m_1} Y^{\m_2}\s^{-1}) \label{eq:GIop}
\end{equation}
There are a large number of elements $\a \in S_n$ that give the same
operator for a given $\{\L,M,a\}$.  We want to describe this
degeneracy.

$\a$ is invariant under its symmetry group $\a = \rho^{-1} \a
\rho$ for $\rho \in \Sym(\a)$.  We can rotate $\s^{-1}$ around the
trace in \eqref{eq:GIop} to get $\tr(\s^{-1}\rho^{-1}\a\rho\s X^{\m_1}
Y^{\m_2})$ and then redefine the sum over $\s$ using $\tau = \rho \s$.
We get a state that is $D^\L(\rho^{-1})$ times the original one.  But
we haven't actually altered the operator since $\a = \rho^{-1} \a
\rho$ so there must be an equivalence class on both $\a$ and the $S_n$
state $a$
\begin{equation}
  (\a,a) \sim \left(\rho^{-1}\a\rho, \sum_a D^{\L}_{ca}(\rho^{-1})\;\; a\right) \quad \rho \in \Sym(\a)
\end{equation}
To remove this redundancy we will back-track and see how the branching
coefficient $B_{b\b}^{\vec{\m}}$ deals with a similar redundancy
\cite{0711.0176}.

\subsubsection{Revisiting the branching coefficient}

The canonical choice of ordering of $X^{\otimes \m_1}\otimes
Y^{\otimes \m_2}$ before we symmetrise in \eqref{SWstate} remains
invariant under $\rho \; X^{\otimes \m_1}\otimes Y^{\otimes \m_2}\;
\rho^{-1}$ for $\rho \in S_{\m_1}\times S_{\m_2}$.  For $\s$ this is a
symmetry on the right $\s \to \s \rho$ for $\rho\in S_{\m_1}\times
S_{\m_2}$ that leaves the state invariant.  To remove this redundancy
we decompose the projector $P = \frac{1}{|S_{\m_1}\times S_{\m_2}|}
\sum_{\rho \in S_{\m_1}\times S_{\m_2}} \;\rho$ into branching
coefficients $B_{b\b}^{\vec{\m}}$
\begin{equation}
  \frac{1}{|S_{\m_1}\times S_{\m_2}|} \sum_{\rho \in S_{\m_1}\times S_{\m_2}} D^\L_{ab}(\rho) = \sum_\b B_{a\b}^{\vec{\m}} B_{b\b}^{\vec{\m}}\label{eq:branchproj}
\end{equation}
This can be understood more intuitively as picking out the trivial
representation ${\bf 1}$ of $S_{\m_1}\times S_{\m_2}$ when we break
$\L$ down into irreps of this subgroup.  \eqref{eq:branchproj} then
becomes
\begin{equation}
  \mbra{\L,a}P\mket{\L,b} = \sum_{\b} \mbraket{\L,a}{\L\to{\bf 1}\textrm{ of }S_{\m_1}\times S_{\m_2};\b}\mbraket{\L\to{\bf 1}\textrm{ of }S_{\m_1}\times S_{\m_2};\b}{\L,b}
\end{equation}
We choose (non-uniquely) the $B_{b\b}^{\vec{\m}}$ to be orthogonal in
the sense
\begin{equation}
  \sum_a  B_{a\b}^{\vec{\m}} B_{a\b'}^{\vec{\m}} = \delta_{\b\b'}\label{branchorthog}
\end{equation}
$\b$ labels the degeneracy of the appearance of the trivial ${\bf 1}$
representation when $\L$ is broken down into irreps of $S_{\m_1}\times
S_{\m_2}$.  The number of values for $\b$ is the Kostka number
$K(\vec{\m},\L)$ given by
\begin{equation}
  K(\vec{\m},\L) \equiv g([\m_1],[\m_2];\L) = \frac{1}{|S_{\m_1}\times S_{\m_2}|} \sum_{\rho \in S_{\m_1}\times S_{\m_2}} \chi_\L(\rho) \label{eq:kostka}
\end{equation}
Note that it is just the trace of the projecting matrix in
\eqref{eq:branchproj}.  The Kostka number $K(\vec{\m},\L)$ counts the
number of ways the field content $\vec{\m}$ can fit into a
semi-standard tableau for $\L$.  It can be written using the
Littlewood-Richardson coefficient $g([\m_1],[\m_2];\L)$ for the
appearance of $\L$ in the $U(2)$ tensor product $[\m_1]\circ[\m_2]$.
The different semi-standard tableaux label the states of the $U(2)$
rep $\L$, so $\{\vec{\m},\b\}$ completely label them too.

To check that the operator \eqref{SWstate} has the required symmetry
on the right, multiply $\s$ by $\rho\in S_{\m_1}\times S_{\m_2}$ to get
\begin{equation}
  B_{ b\b}^{\vec{\m}}\; D^{\L}_{ab}(\s\rho) = D^{\L}_{bc}(\rho)B_{ b\b}^{\vec{\m}}\; D^{\L}_{ac}(\s) = B_{ c\b}^{\vec{\m}}\; D^{\L}_{ac}(\s)
\end{equation}
This follows by using \eqref{branchorthog} to prove that $B_{
  b\b}^{\vec{\m}} = \sum_{\b'} B_{ b\b'}^{\vec{\m}}\; B_{
  d\b'}^{\vec{\m}}\; B_{ d\b}^{\vec{\m}}$ and then using
\eqref{eq:branchproj} to absorb $D^{\L}_{bc}(\rho)$.  The $B_{
  b\b}^{\vec{\m}}$ are an orthogonal basis for the eigenspace of the
projector \eqref{eq:branchproj} with eigenvalue 1.

\subsubsection{Removing trace description redundancy}

$\a$ is a way of describing the trace structure of the operator. For
each conjugacy class of $S_n$ choose $\a$ to be the canonical
permutation. The presence of $\a$ in the gauge-invariant operator
\eqref{eq:GIop} with $\s^{-1} \a \s$ induces a symmetry, this time on
the left $\s \to \rho\s$ where $\rho$ is in the symmetry group
$\Sym(\a)$ of $\a$ so that $\a = \rho^{-1} \a\rho$.  To remove this
redundancy we proceed in exactly the same way using a coefficient
$S^{\a}_{a\g}$ for $S_n \to \Sym(\a)$ that decomposes the projector
\begin{equation}
  \frac{1}{|\Sym(\a)|} \sum_{\rho \in \Sym(\a)} D^\L_{ab}(\rho) = \sum_{\g}  S_{a\g}^\a S_{b\g}^\a \label{eq:tracedeomcp}
\end{equation}
We choose the  $S_{a\g}^\a$ to be orthogonal in the sense
\begin{equation}
  \sum_a  S_{a\g}^\a S_{a\g'}^\a = \delta_{\g\g'}
\end{equation}
This orthogonal choice is not unique.  Examples of the $S_{a\g}^\a$
for $\L=[2,2]$ and for $\L=[4,2]$ with trace structure $\a = [4,2]$
are computed in Appendix Section \ref{sec:trexamples}.  Contracting
$S_{a\g}^\a$ with the degenerate operator \eqref{eq:GIop} gives us our
final non-degenerate operator
\begin{equation}
\boxed{\ket{\L,M;\a,\g} = \frac{1}{n!} \sum_{\s \in S_n} S^{\a}_{a\g} \;B_{b\b}^{\vec{\m}}\; D^{\L}_{ab}(\s) \;\tr(\a\s X^{\m_1} Y^{\m_2}\s^{-1})} \label{tracegambas}
\end{equation}
This is the crucial result of this paper.  Note that we pick a single
canonical $\a$ from each conjugacy class of $S_n$, so that $\a$ is
just a partition of $n$.  The $\g$ label runs over the
elements of each conjugacy class (i.e. trace structure) which are
compatible with the symmetry imposed by $\L$. 

In Appendix Section
\ref{sec:invert} it is checked that the basis defined by
\eqref{tracegambas} is complete, in that any multi-trace operator of
$X$'s and $Y$'s can be built out of a linear combination of this
basis.  We check below that these operators give the correct counting.

The number of values for $\g$ is  given by the trace of the projecting matrix
\begin{equation}
  \boxed{S(\a,\L) = \frac{1}{|\Sym(\a)|} \sum_{\rho \in \Sym(\a)} \chi_\L(\rho)} \label{eq:tracecount}
\end{equation}
Given the symmetry imposed by $\L$ this number tells us how many
operators there are with trace structure $\a$. It gives an analytic
expression for the examples computed in \cite{0405057}.  The partition
function for global group $U(2)$ for $N\to \infty$ can thus be
decomposed into $U(2)$ characters $\chi_{\L}(x,y)$
\begin{equation}
  Z(x,y) = \prod_{i=1}^\infty\frac{1}{1-(x^i+y^i)} = \sum_{n}\sum_{\L \in P(n,2)} \sum_{\a\in P(n)}\; S(\a,\L) \;\; \chi_{\L}(x,y)
\end{equation}
This result is checked against known results \cite{0704.1038,0711.0176} in
Appendix Section \ref{sec:counting}.

Note that the $\s \in S_n$ sum in the final operator
\eqref{tracegambas} is invariant both on the left by $\Sym(\a)$ and on
the right by $S_{\m_1} \times S_{\m_2}$ so we could just sum over the
double coset of $S_n$
\begin{equation}
  \Sym(\a)\, \backslash\, S_n\, /\, S_{\m_1} \times S_{\m_2}
\end{equation}

\subsection{Inner product and two-point function}\label{sec:geninner}

Before we give the inner product and the two-point function for these
operators, we need to define the conjugate with care.  The order of
the fields in each trace are reversed so that the bra is defined with
the inverse $\a^{-1}$ of the canonical permutation in each conjugacy class\footnote{NB: The symmetry group of $\a$ and its inverse are the same $\Sym(\a) = \Sym(\a^{-1})$.}
\begin{equation}
  \bra{\L,M;\a,\g} = \frac{1}{n!} \sum_{\s \in S_n} S^{\a}_{a\g}\; B_{b\b}^{\vec{\m}}\; D^{\L}_{ab}(\s)\; \tr(\a^{-1}\s X^{\dagger \m_1} Y^{\dagger\m_2}\s^{-1})
\end{equation}
The inner product for the gauge-invariant operators defined by
\eqref{tracegambas} then corresponds to the planar two-point function
function, which is now diagonal
\begin{equation}
\boxed{  \braket{\L',M';\a',\g'}{\L,M;\a,\g} =\delta^{\L\L'}\delta^{MM'} \delta^{\a\a'}\delta^{\g\g'}\;\; \frac{| \Sym(\a) | | S_{\m_1} \times S_{\m_2} |}{n!d_\L}} \label{eq:innergen}
\end{equation}
The matrix elements for the cut-and-join operators are
\begin{align}
 \bra{\L',M';\a',\g'} \S_{C} \ket{\L,M;\a,\g} =\delta^{\L\L'}\delta^{MM'}\frac{|S_{\vec{\m}}|}{n!d_\L} \sum_{\tau \in  S_n} \delta(\S_{C} \a  = \tau^{-1}\a^{\prime}\tau)\;\; S_{a\g}^\a \;S_{a'\g'}^{\a'}\;\; D_{a'a}^{\L}(\tau)\label{eq:gencutjoin}
\end{align}
The canonical planar three-point function between a single-trace state $\a = [n]$ and a double-trace state $\a=[n_1,n_2]$ can then be computed
\begin{align}
 \bra{\L',M';\a'=[n_1,n_2],\g'} \,\S_{[2]}\, \ket{\L,M;\a=[n],\g}
\end{align}
This is easier to understand if we split the left state into two
single-trace states
\begin{align}
\boxed{ \left(\bra{\L'_1,M'_1;[n_1],\g'_1}\otimes  \bra{\L'_2,M'_2;[n_2],\g'_2}\right) \,\S_{[2]}\, \ket{\L,M;\a=[n],\g} = C^{\L_1'\circ\L_2'=\L}_{M_1'\,\,M_2'\,\,M}\;\; f(\g'_1,\g'_2;\g)} \label{eq:messsy3pt}
\end{align}
We get a Clebsch-Gordan coefficient coupling the $U(2)$ states and a
function $f$ of the $\g$'s.  See Appendix Section \ref{sec:3ptvertex}
for more details.

The full non-planar free two-point function follows from the cut-and-join operators \eqref{eq:gencutjoin}
\begin{align}
\corr{\cO^\dagger[\L',M';\a',\g']\cO[\L,M;\a,\g]} & =\delta^{\L\L'}\delta^{MM'}  \frac{|S_{\vec{\m}}|}{n!d_\L} \sum_{\tau \in S_n} S^{\a}_{a\g}S^{\a'}_{a'\g'} D^{\L}_{a'a}(\tau) N^{C(\tau^{-1} \a \tau\a^{\prime -1})} \nn \\
& = \delta^{\L\L'}\delta^{MM'}\;N^n \,\bra{\L,M;\a',\g'}\, \O_n \,\ket{\L,M;\a,\g} 
\end{align}
This is only diagonal in the global symmetry group $U(2)$ labels.  As
in all previous cases, the appearance of $\O_n$ means that the
non-planar correlation functions factorise into planar three-point
functions when $n$ is large as in equation \eqref{eq:expvertex}.
Because the notation above is general the extension to all other
unitary groups $U(K)$ is trivial.  Thus the structure observed in the
half-BPS sector, including the large $n$ factorisation, is universal
for the whole of $\cN=4$ SYM when the coupling is turned off.

\subsection{Relation to finite $N$ bases}

To take account of the Stringy Exclusion Principle for operators with
$\D \geq N$ the traces can be reorganised into representations $R$ of
$U(N)$ just as for the Schur polynomials of half-BPS operators in
Section \ref{sec:schurpoly}.

First consider the basic operator $\ket{\vec{\m},\a}$ from
\eqref{eq:basiop}, before we've organised into $U(2)$ reps.
Unlike the Schur polynomials, this is not a class function of $\a$, so
we cannot change basis using the class-invariant $S_n$ character; we
must use the more general matrix representation $ D^R_{pq}(\a)$ of
$S_n$
\begin{equation}
  \ket{\vec{\m};R,p,q} = \frac{1}{n!} \sum_{\a} D^R_{pq}(\a) \ket{\vec{\m};\a}
\end{equation}
The floating $p,q$ state indices of $V_R^{S_n}$ will be dealt with
later.  To work out the action of the cut-and-join operators on these
states, first note that as a consequence of Schur's lemma the central
elements $\S_{C}$ act multiplicatively
\begin{equation}
  D_{pq}^R(\S_{C}\a) =  \frac{\chi_R(\S_{C})}{d_R}D_{pq}^R(\a)
\end{equation}
This means that the $U(N)$-organised trace is an eigenvector of the
$\S_C$
\begin{equation}
  \S_{C} \ket{\vec{\m};R,p,q} = \frac{\chi_R(\S_{C})}{d_R} \ket{\vec{\m};R,p,q} \label{eq:cutjoinR}
\end{equation}

One way to absorb the $p,q$ indices is with a particular type of
branching coefficient to build the `restricted' Schur polynomials
\cite{0411205,ggstringsattach,0801.2061}.  Another way \cite{0711.0176} is to
organise the ${\bf X}^{\vec{\m}}$ into representations $\{\L,M,a\}$ of
$U(2) \times S_n$ as we did in Section \ref{sec:organrep} and contract
the remaining $S_n$ indices with an $S_n$ $3j$ Clebsch-Gordan
coefficient $ S^{\hat\tau ,}{}^{ \L }_{ a}\;{}^{R}_{p }\;{}^{R}_{q}$
\begin{align}
     \ket{ \L,M; R, \hat\tau}= \frac{1}{n!}\sum_{\a\in S_n}
 S^{\hat\tau ,}{}^{ \L }_{ a}\;{}^{R}_{p
     }\;{}^{R}_{q}\;\; D_{pq}^R(\a)\;\; \ket{\L,M,a;\a}
\end{align}
The relation to the $\ket{\L,M;\a,\g}$ trace basis is given in
Appendix Section \ref{sec:gamtotau}.  Just like the Schur polynomials
they are diagonal in the planar inner product
\begin{equation}
  \braket{\L',M',R',\hat\tau'}{\L,M,R,\hat\tau} = \delta_{\L\L'} \delta_{MM'} \delta_{RR'} \delta_{\hat\tau\hat\tau'}  \frac{|H_\m|}{n! d_R}
\end{equation}
From equation \eqref{eq:cutjoinR} they are eigenstates of the
cut-and-join operators
\begin{equation}
  \S_{C} \ket{\L,M,R,\hat\tau} = \frac{\chi_R(\S_{C})}{d_R} \ket{\L,M,R,\hat\tau} \label{eq:schurmultsd}
\end{equation}
The commuting $\S_C$ distinguish the $U(N)$ representations $R$
by their different eigenvalues. They are linear combinations of the
$U(N)$ Casimirs.

As a conseqence of equation \eqref{eq:schurmultsd} the full non-planar
correlation function is also diagonal
\begin{equation}
  N^n \bra{\L',M',R',\hat\tau'}\O_n \ket{\L,M,R,\hat\tau} = \delta_{\L\L'} \delta_{MM'} \delta_{RR'} \delta_{\hat\tau\hat\tau'} \frac{|H_\m|}{n! d_R^2}\Dim R
\end{equation}

\subsection{Trace operators for reps of general groups}\label{sec:gengroup}

In equation \eqref{eq:SWdecomp} $n$ copies of the fundamental
representation $V_{\bf 2}$ of $U(2)$ were decomposed into general
irreps of $U(2) \times S_n$.  A precise formula for the states in
these irreps was given in equation \eqref{SWstate}.  Here we will give
a more abstract description of these decompositions using
Clebsch-Gordan coefficients.  Label the fundamental fields of $U(2)$ by $W_1 = X$ and $W_2 = Y$.  Define the Clebsch-Gordan coefficient for
\begin{equation}
    V_{\bf 2}^{\otimes n} \to V_{\L}^{U(2)} \otimes V_{\L}^{S_n} 
\end{equation}
by $C^{\vec{m}}_{\L,M,a}$ so that the state in \eqref{SWstate} is
\begin{equation}
\mket{\L,M,a}  =  \sum_{\vec{m}}C^{\vec{m}}_{\L,M,a}  \; W_{m_1} \otimes W_{m_2} \otimes \cdots \otimes W_{m_n} \label{eq:covop}
\end{equation}

For a group $G$ other than $U(K)$ the decomposition of $n$ copies of
one of its representations $V_F$ into irreps $\L \otimes \l$ of $G
\times S_n$ is not necessarily multiplicity-free as it was in
\eqref{eq:SWdecomp}.  An irrep $\L \otimes \l$ generically appears
with an integer multiplicity $\mult(\L,\l)$
\begin{equation}
  V_F^{\otimes n} = \bigoplus_{\L,\l}\;\; \mult(\L,\l)\;\;\;V_\L^G \otimes V_\l^{S_n}  \label{eq:decomp}
\end{equation}
In the states for this decomposition this multiplicity is labelled by
$\tau$
\begin{equation}
\mket{\L,M,\l,a,\tau}  =  \sum_{\vec{m}}C^{\vec{m}}_{\L,M,\l,a,\tau}  \; W_{m_1} \otimes W_{m_2} \otimes \cdots \otimes W_{m_n}  \label{eq:gencovop}
\end{equation}
$M$ is still the state of the $G$ irrep $\L$; $a$ is the $S_n$ state
of $\l$.  The Clebsch-Gordan coefficient is defined to satisfy various
orthogonality conditions; see \cite{0806.1911} for more details.

This decomposition was outlined in \cite{0711.0176} for $G=
U(K_1|K_2)$ and in \cite{0806.1911} for $G=SO(6)$, the symmetry for
the six real scalars of $\cN=4$ SYM, and $G = SO(2,4)$, where $V_F$ is
now a single scalar with all possible onshell combinations of the four
derivatives (for one derivative and $G=SL(2,\R)$ this was done in
detail).  Extending these results to $PSU(2,2|4)$ where $V_F$ is the
singleton representation is tricky because of shortening conditions
\cite{Dobrev:1985qv,0209056}, however it is done for $n=2$ in \cite{0105001}
and for $n=3$ in \cite{0405057} (the higher spin YT-pletons in the
latter are our $\l$).  For the inner product use the spin bit metric
from \cite{0510264}, see \cite{0806.1911} for more details.

To get the trace operators with trace structure $\a$ ($\a$ is the
canonical choice within its conjugacy class) we contract the $S_n$
state $a$ of $V_{\l}^{S_n}$ with the same coefficient $S_{a\g}^{\a}$
that we used before for $U(2)$ in equation \eqref{tracegambas} to get
\begin{equation}
\boxed{\ket{\L,M,\l,\tau;\a,\g}  =  \sum_{\vec{m}}\;\; S_{a\g}^\a\;\;C^{\vec{m}}_{\L,M,\l,a,\tau}  \; \tr(\a\;W_{m_1}\,W_{m_2} \cdots W_{m_n} )}
\end{equation}
The counting for a representation $\L$ and trace structure $\a$ must now take account of the multiplicity in the decomposition \eqref{eq:decomp}
\begin{equation}
  S(\a,\L) = \sum_{\l \in P(n)} \mult(\L,\l) \;\; \frac{1}{|\Sym(\a)|} \sum_{\rho \in \Sym(\a)} \chi_\l(\rho)
\end{equation}
The orthogonality in the planar inner product and the non-planar
structure of the free theory through $\O_n$ follow exactly as for
$U(K)$.  Essentially the trace only sees how the operator is symmetrised
through the $S_n$ irrep $\l$.

\section{Application to the chiral ring at weak coupling}\label{sec:weak}

In this section we present an application of the technology developed
in Section \ref{sec:general}.  We construct the genuine quarter-BPS
operators at weak coupling, including all their $\frac{1}{N}$
corrections.  Crucial to this result is the understanding of how to
enumerate the different trace structures for each $U(2)$
representation and then how to ensure orthogonality in the non-planar
two-point function to the newly descendant anomalous operators at weak
coupling.  The extension to eighth-BPS states in $SU(3|2)$
representations should be possible using the Schur-Weyl technology for
supergroups developed in \cite{0711.0176}.

\subsection{Introduction}

At one-loop the spectrum of anomalous dimensions and the mixing in the
two-point function are connected via the one-loop dilatation operator
\cite{0205089,0208178,0208231,0209263,0212269}
\begin{equation}
  \D = \tr([X,Y][\tilde X, \tilde Y])
\end{equation}
where $\tilde X^i_j \equiv \frac{\d}{\d X^j_i}$.  $\D$ can act on
operators to give the matrix of anomalous dimensions or it can be
inserted into the two-point function as an effective vertex for the
one-loop correction.

If we diagonalise the action of $\D$ on the space of $U(2)$
multi-trace operators we find that the space splits into those
operators with non-trivial anomalous dimensions and those operators
with no correction to their dimension at one-loop.  Because $\D$
always inserts a commutator $[X,Y]$ into the trace on which it acts,
the non-trivial eigenstates always have commutators within traces.  At
one-loop these operators have in fact become descendants in long
representations of other operators (i.e. they are no longer
highest-weight states).  For example the operator
\begin{equation}
  \tr([X,Y][X,Y]) \label{eq:L22desc}
\end{equation}
becomes a descendant of the Konishi operator at one-loop. This
discontinuous change in the spectrum from zero coupling to one-loop can
also be seen from the action of the supercharge on the fermion which
gains an additional term at weak coupling
\begin{equation}
  Q \l \sim F + g[X,Y]
\end{equation}

On the other hand the operators which remain BPS at weak coupling,
with no correction to their dimension, are given at large $N$ by those
multitrace operators built assuming that the fields commute within
each trace. In these symmetrised traces we sum over all orders of the
fields within the trace. This process removes all commutators inside
traces, but commutators can still cross between two different
traces. These operators built from symmetrised traces are part of the
\emph{chiral ring} \cite{0211170}.  For $\L = \tyng(2,2)$ there are
two such operators, $\tr(\Phi_r \Phi_s)\tr(\Phi^r)\tr(\Phi^s)$ and
$\tr(\Phi_r \Phi_s)\tr(\Phi^r\Phi^s)$ where $\Phi^1 = X$, $\Phi^2 = Y$
and $\Phi_r \Phi^r = \e_{rs} \Phi^r \Phi^s = [X,Y]$.

For $N$ finite this does not completely describe the BPS operators.
The BPS operators must be annihilated by the dilatation operator and
be orthogonal in the full non-planar two-point function to the
anomalous descendant operators (these two conditions are `if and only
if' since on general grounds operators with different dimensions are
orthogonal in the two-point function of a CFT).  This requires
$\frac{1}{N}$ corrections to be added to the operators built from
symmetrised traces \cite{0109064,0301104}.  For example for the $\L =
\tyng(2,2)$ case we must add the descendant operator from
\eqref{eq:L22desc} to $\tr(\Phi_r \Phi_s)\tr(\Phi^r\Phi^s)$ to get the
genuine BPS operator
\begin{equation}
  \tr(\Phi_r
\Phi_s)\tr(\Phi^r\Phi^s) + \frac{2}{N} \tr(\Phi_r\Phi^r\Phi_s\Phi^s)\label{eq:examplegenuinBPS}
\end{equation}
The second term is suggestive of a joining  accompanied by
$\frac{1}{N}$.

The goal of this section is to capture these non-planar corrections in
a precise analytic way, building on work in \cite{0301104,0711.0176},
and connect the analysis to the cut-and-join operators constructed
above.  This is only possible with the technology from Section
\ref{sec:general}, because we need to know exactly how the $U(2)$
representation fits into each trace.

\subsection{The dual basis}

Take a representation and state $\{\L,M\}$ of $U(2)$ and split the
compatible trace structures $\{ \a, \g\}$ into two groups ($\a$ is the
canonical permutation for a particular conjugacy class of $S_n$ and
$\g$ labels its multiplicity for this $U(2)$ representation)
\begin{itemize}
\item Identify $\{\a^d,\g^d\}$ exactly with those operators with
  commutators inside traces.  These operators may be linear
  combinations of the original $\{\a,\g\}$ identified in the
  tree-level section.  They coincide with the image of $\D$ and each
  of them is a descendant at one-loop.  Because they all appear in the
  image of $\D$, they have anomalous dimensions.  However in this
  basis they may not be diagonal under $\D$: the $\{\a^d,\g^d\}$ could
  be a linear combination of the non-trivial eigenstates of
  $\D$.\footnote{See Table \ref{tab:42estates} in Appendix Section
    \ref{sec:42} for examples of eigenstates that mix
    trace-structures for $\L=[4,2]$.}
  \item Identify the remainder by $\{\a^c,\g^c\}$ such that they are
    orthogonal to all the $\{\a^d,\g^d\}$ in the \emph{planar} inner
    product.  These will give the leading terms for the genuine
    quarter-BPS operators, which are known on general grounds to be
    built from symmetrised traces.
    This is a much simpler job than insisting on orthogonality in the
    non-planar two-point function: if $\a^c$ has a different trace
    structure to each descendant in $\{\a^d,\g^d\}$ then we are done
    because different trace structures are automatically orthogonal in
    the planar inner product \eqref{eq:innergen}.  If however for a
    given trace structure $\a$ some operators contain commutators
    within traces and some don't, then we must take care to guarantee
    orthogonality.  An example of this type for $\L=[4,2]$ is computed
    in Appendix Section \ref{sec:42}.
\end{itemize}

Thus under the dilatation operator the $\{\a^d, \g^d\}$ furnish the
image of $\D$
\begin{equation}
  \D\ket{\L,M;\a,\g} =\sum_{\{\a^d,\g^d\}} D_{\{\a,\g\};\{\a^d,\g^d\}} \ket{\L,M;\a^{d},\g^{d}}
\end{equation}
where $D$ is a non-diagonal matrix of anomalous dimensions.  The $\{\a^c,\g^c\}$ satisfy  the condition
\begin{equation}
  \braket{\L,M;\a^c,\g^c}{\L,M;\a^d,\g^d} = 0\quad \quad \forall \;\;c,d
\end{equation}

We are now ready to define the genuine quarter-BPS operators with all
their $\frac{1}{N}$ corrections:
\begin{equation}
  \boxed{\ket{\L,M;\a^c,\g^c; \perp} \equiv \O_n^{-1}\;\ket{\L,M;\a^c,\g^c}} \label{eq:quarterdef}
\end{equation}
Note that $\O_n^{-1}$ only exists if $N \geq n$; its first few terms
are given in Appendix Section \ref{sec:on}.  In previous work
\cite{0703202,0711.0176} such operators have been called a \emph{dual
  basis} because they are dual to the original trace basis in the
two-point function.  The connection to these papers is spelt out
below.

We will now show that these operators satisfy two important
properties:
\begin{itemize}
\item They are orthogonal to the descendant non-trivial eigenstates in
  the full non-planar two-point function at tree level and at one
  loop;
\item They are annihilated by the dilatation operator.
\end{itemize}
That the second property follows from the first on general CFT grounds
is the basis of the analysis in \cite{0301104,0711.0176}.  The exact
description of the non-planar corrections in Section \ref{sec:general}
for each trace structure allows us to give precise analytic formulae
here.

The orthogonality in the tree-level two-point function follows by definition
\begin{align}
  \corr{\cO^\dagger[\L,M;\a^c,\g^c;\perp]\;\cO[\L,M;\a^d,\g^d]}_{\textrm{tree}}& = \langle \L,M;\a^c,\g^c;\perp| \O_n \ket{\L,M;\a^d,\g^d} \nn \\
& = \langle\L,M;\a^c,\g^c| \O_n^{-1}  \O_n\ket{\L,M;\a^d,\g^d} \nn \\
& = \braket{\L,M;\a^c,\g^c}{\L,M;\a^d,\g^d} \nn \\
& = 0
\end{align}
Orthogonality in the two-point function at one-loop follows
automatically because if we insert $\D$ above only descendants appear
in its image so the same logic follows.

The annihilation of the quarter-BPS operators by $\D$ follows in a
similar way by considering the one-loop two-point function
\begin{equation}
  \corr{\cO^\dagger[\L,M;\a,\g]\;\cO[\L,M;\a^c,\g^c;\perp]}_{\textrm{1-loop}} = \bra{\L,M;\a,\g}\O_n \D \ket{\L,M;\a^c,\g^c;\perp} \label{eq:1lop2pt}
\end{equation}
Since $\O_n$ is non-singular and the bra $\bra{\cdots}$ contains
everything that could possibly be in the image of $\O_n \D$, if this
object vanishes for all trace structures $\{\a,\g\}$ then it will
follow that
\begin{equation}
  \D\ket{\L,M;\a^c,\g^c,\perp} = 0
\end{equation}

By the symmetry of the two-point function, \eqref{eq:1lop2pt} equals
its conjugate
\begin{align}
  \bra{\L,M;\a^c,\g^c;\perp}\O_n \D\ket{\L,M;\a,\g}&  =  \sum_{\{\a^d,\g^d\}}\bra{\L,M;\a^c,\g^c}\O_n^{-1}\;\O_n\; D_{\{\a,\g\};\{\a^d,\g^d\}} \;\ket{\L,M;\a^{d},\g^{d}} \nn \\
& =  \sum_{\{\a^d,\g^d\}}\; D_{\{\a,\g\};\{\a^d,\g^d\}} \;\braket{\L,M;\a^c,\g^c}{\L,M;\a^{d},\g^{d}} \nn \\
&=0
\end{align}
Thus the $\ket{\L,M;\a^c,\g^c,\perp}$ are indeed annihilated by $\D$
and hence have a protected scaling dimension, at least at one-loop.
It is conjectured \cite{0307015} that these operators remain
protected for higher loops too, since the higher order dilatation
operators all contain the common element $[\tilde X,\tilde Y]$, which annihilates the operators constructed here.

\subsection{Further analysis of dual basis using other 1-loop formulations}\label{sec:further}

In this section we sketch a complementary analysis of why the the
dual basis for the $\{ \a^c,\g^c\}$ is annihilated by the dilatation
operator $\D = \tr([X,Y][\tilde X, \tilde Y])$ using two alternative
formulations of the action of $\D$.

$\D$ acts on two sites at a time, cycling through all possible pairs
($1\leq i,j\leq n$) by the product rule.  When it acts on the global
symmetry group part it projects onto the antisymmetric $\tyng(1,1)
\sim [X,Y]$ combination, splitting this off from the $U(2)$
representation $\{\L,M\}$ (this separation is expressed clearly in
\cite{0806.1911}).

It also alters the trace structure.  There are two equivalent ways of
writing the action on the trace permutation $\a$:
\begin{itemize}
\item The first way was first demonstrated in \cite{0404066} using the
  spin bit formalism.  The action of $\D$ is summarised by the left
  action on $\a$ by transpositions $(i\,\a(j))$
  \begin{equation}
    \D \sim \sum_{i\neq j} P_{ij}\;\;\; (i\,\a(j))\, \a \label{eq:spinform}
  \end{equation}
  $P_{ij}$ is the projector onto $[X,Y]$ acting on the fields.  In
  some ways this is similar to the action of $\S_{[2]}$, except that
  $P_{ij}$ is only non-vanishing when a commutator is present and the
  transposition $(i\,\a(j))$ requires knowledge of $\a$ beforehand.
  When $\a(j)=i$ then $(i\,\a(j))\equiv N$; in this case the two sites
  of the commutator are adjacent in the trace.
\item To avoid the dependence of $\D$ on $\a$ and derive
  explicit expressions for non-planar one-loop mixing matrices this
  action can also be written by introducing an extra $(n\pd1)$th
  permutation index and both pre- and post-multiplying $\a$
  \cite{0801.2094}
  \begin{equation}
    \D \sim \sum_{i\neq j} P_{ij} \;\;\;(i\,n\pd1)\,\a\,(j\,n\pd1)
  \end{equation}
  For the finite $N$ bases this gives the one-loop two-point function
  as an expansion of dimensions of $U(N)$ representations with $n\pd1$
  boxes.  Operators with $n$-box $U(N)$ representations only mix if
  they are related by moving a single box in the Young diagram.
\end{itemize}
In Appendix Section \ref{sec:1lpequiv} we show the equivalence of
these two formulations; in the remainder of this section we use the
first.

We want to show that the dilatation operator of the form
\begin{equation}
  \Delta = \sum_{\{i,j\}\sim [X,Y]}\;\; (i\,\a(j)) \label{eq:roughdil}
\end{equation}
annihilates the quarter-BPS operators defined in
\eqref{eq:quarterdef}.  The latter have a trace structure
\begin{equation}
  \O_n^{-1} \a = \left[1 - \frac{1}{N} \S_{[2]} + \cO\left( \frac{1}{N^2} \right) \right]\; \a
\end{equation}
where in $\a$ the sites of each commutator $i$ and $j$ are always in
different cycles (i.e. the operator's leading term is the product of
symmetrised traces).  More terms of $\O_n^{-1}$ are given in Section
\ref{sec:on}.

For the leading term, $\a$, the dilatation operator $\D$ will join the
two cycles which contain $i$ and $j$ so that they are in the same
cycle (and hence describe a descendant operator).

When $\S_{[2]}$ joins traces in $\a$ we get operators which are
descendant ($i$ and $j$ in same cycle).  Under $\D$ this gets a factor
$N$, cancelling the $\frac{1}{N}$ and then cancelling against the
descendant from the leading term $\D\; \a$.

If $\S_{[2]}$ splits traces in $\a$ then we still get something that
is in the chiral ring.  Acting with $\D$ we get a descendant at order
$\frac{1}{N}$ which must be cancelled by a descendant from
$\frac{1}{N^2}\left( \S_{[3]}+ \S_{[2,2]}\right)\a$ that gains a
factor of $N$ from $\Delta$.

This cancellation continues at each order in $\frac{1}{N}$ until there
is nothing left.  Thus the quarter-BPS operator is indeed in the
kernel of $\D$.

\subsection{Dual basis for $N < n$ and $SU(N)$}

At finite $N$, in particular when $N$ is less than the length of the
operator $N < n$, we cannot define $\O_n^{-1}$.  Fortunately by
expanding $\O_n^{-1}$ in terms of $U(N)$ dimensions using
\eqref{eq:OfromDim} we can define the dual basis by truncating the sum over $R$ in \eqref{eq:OfromDim} to those Young diagrams with $N$ or fewer rows
\begin{equation*}
  \ket{\L,M;\a^c,\g^c; \perp} = \frac{N^n}{(n!)^2} \sum_{\s \in S_n}
  \sum_{R\in P(n,N)} \frac{d_R^2}{\dim R}\; \chi_R(\s)\;\s
\; \ket{\L,M;\a^c, \g^c}
\end{equation*}
This continues to be dual to the trace basis in the non-planar
two-point function for $N < n$.  It is the generalisation to
arbitrary representation $\L$ of the dual basis constructed in
\cite{0703202} for the half-BPS operators.  If we remove the
restriction on $R\in P(n,N)$ that there are no more than $N$ rows in
the Young diagram then this expression is identical to
\eqref{eq:quarterdef}.

To extend the trace basis to $SU(N)$, where the traces of the single
adjoint fields vanish, just drop the $\a$ with 1-cycles in them.  It
turns out that the dual basis as defined in \eqref{eq:quarterdef} for
such $\a$ is also dual to the trace basis in the $SU(N)$ two-point
function \cite{0703202}, so these results for the quarter-BPS
operators should follow through for the more physical $SU(N)$ gauge
group.

\section{Counting the chiral ring using the Weyl group $S_N$}\label{sec:weyl}

In this section we characterise the chiral ring of 4d $\cN=4$ super
Yang-Mills at finite $N$ in terms of functions of the eigenvalues of the matrix fields and representations of the subgroup
$S_N$ of the gauge group $U(N)$.  The number of these
operators matches the finite $N$ partition functions computed in Dolan
\cite{0704.1038} and furthermore provides a counting of chiral ring
operators for each representation of the global symmetry group $G$.
$G$ is always a subgroup of $SU(3|2)$ for the chiral ring of $\cN=4$
SYM, corresponding to $\tfrac{1}{8}$th-BPS operators, but because
these methods are applicable to any eigenvalue system we leave the
group general.

\subsection{Invariant functions of eigenvalues}

In previous sections we considered gauge-invariant operators built out
of generic matrices transforming in the adjoint of $U(N)$.  Here we
consider the chiral ring, a subset of operators built out of commuting
matrices.  These are functions only of the eigenvalues, since the
matrices are simultaneously diagonalisable.  These symmetric functions
of eigenvalues are organised by irreps of the $S_N$ which permutes the
eigenvalues and the $S_n$ which permutes tensor products of
fundamental fields.

In Section \ref{sec:gengroup} we organised tensor products of the
fundamental fields $V_F^{\otimes n}$ for a global symmetry group $G$
into representations $\L \times \l$ of $G \times S_n$ 
\begin{equation}
\mket{\L,M,\l,a,\tau}  =  \sum_{\vec{m}}C^{\vec{m}}_{\L,M,\l,a,\tau}  \; W_{m_1} \otimes W_{m_2} \otimes \cdots \otimes W_{m_n} \label{eq:covop2}
\end{equation}
$M$ is the state of the representation $\L$ of $G$, $a$ is the state of
$\l$ of $S_n$ and $\tau$ labels the multiplicity with which $\L \times
\l$ appear in $V_F^{\otimes n}$.

Now consider the eigenvalues of these fundamental fields $w^e_m$ where
$e \in \{1, 2, \dots N\}$.  The subgroup of the gauge group $U(N)$
which acts on these eigenvalues is $S_N$, the symmetric group which
permutes the eigenvalues.  The eigenvalues are in the \emph{natural
  representation} $V^{S_N}_{\nat}$ of $S_N$, the $N$-dimensional
representation where $S_N$ acts by just permuting the elements.\footnote{This representation of $S_N$ is also known as the permutation representation.}  $V^{S_N}_{\nat}$ is reducible
\begin{equation}
  V^{S_N}_{\nat} = V_{[N]}^{S_N} \oplus V^{S_N}_{[N-1,1]}
\end{equation}
The trivial representation $V_{[N]}^{S_N}$ is the sum of the
eigenvalues, which is invariant under $S_N$; $V^{S_N}_{[N-1,1]}$,
referred to as the `standard' representation in \cite{FultonHarris}
and the `hook' representation in \cite{0806.1911}, represents the
$N-1$ differences of the eigenvalues.\footnote{This split between
  differences and the sum was used in \cite{0806.1911} to distinguish
  highest weight states from descendants in the tensor products of
  $SL(2)$ spin $-\frac{1}{2}$ representations.}  To change the gauge group from $U(N)$ to $SU(N)$, for
which $S_N$ is the Weyl group, just substitute $V^{S_N}_{\nat}$ with
$V^{S_N}_{[N-1,1]}$ in any of the following discussion.

We can use Schur-Weyl duality on the $n$-tensor product of the natural
representation of $S_N$ to decompose it into representations $K \times
\k$ of $S_N \times S_n$
\begin{equation}
\left(  V^{S_N}_{\nat} \right)^{\otimes n} = \bigoplus_{K\vdash N,\k\vdash n} \mult(K,\k)\;\;V_{K}^{S_N} \otimes V_{\k}^{S_n}  \label{eq:SNSW}
\end{equation}
$K\vdash N$, which is equivalent to $K \in P(N)$ in our previous
notation, means that $K$ is a partition of $N$.  The multiplicity
$\mult(K,\k)$ for the appearance of $K \times \k$ is labelled with
$\check \tau$ in the Clebsch-Gordan coefficent
$C^{\vec{e}}_{K,M_K,\k,a_{\k},\check \tau}$ for the decomposition
\eqref{eq:SNSW}.  Projecting onto $K \times \k$ in $(
V^{S_N}_{\nat})^{\otimes n}$ gives a formula for $\mult(K,\k)$, see
equation \eqref{eq:multexpan} in Appendix Section
\ref{sec:CRproof}.\footnote{The multiplicity-free Schur-Weyl dual of
  $S_N$ is the maximal algebra in the space of endomorphisms of $(
  V^{S_N}_{\nat} )^{\otimes n}$ that commutes with $S_N$: it is known
  as the partition algebra $P_n(N)$ \cite{martin,jones}.  The
  symmetric group algebra is a subalgebra of $P_n(N)$ via the Brauer
  algebra $B_n(N)$ (which is the Schur-Weyl dual of $O(N)$), $\C S_n
  \subset B_n(N) \subset P_n(N)$, which mirrors the fact that $U(N)
  \supset O(N) \supset S_N$.  As the group gets smaller, the commuting
  algebra grows.}

We can thus map the space of eigenvalues $(V_F^G \otimes
V^{S_N}_{\nat} )^{\otimes n}$ to the linear combinations
\begin{equation}
  C^{\vec{m}}_{\L,M,\l,a,\tau} \;\;  C^{\vec{e}}_{K,M_K,\k,a_{\k},\check \tau} \;\;\; w^{{e}_1}_{{m}_1} \; w^{{e}_2}_{{m}_2}  \; \cdots \; w^{{e}_n}_{{m}_n} 
\end{equation}
For the operators of the chiral ring, we know that they are invariant
under the $S_N$ that permutes the eigenvalues (this is the remnant of
the $U(N)$ gauge invariance that survives for the eigenvalues).  This
means $K$ is the trivial representation of $S_N$, $[N]$.  Furthermore
the final operators should be an overall $S_n$ invariant too, because
the eigenvalues are commuting bosons.  This forces $\l = \k$ and
requires us to sum over the $S_n$ states $a = a_\k$.  Thus we get
the chiral ring as functions of eigenvalues
\begin{equation}
\boxed{ \ket{\L,M,\l,\tau,\check \tau} = \sum_a  C^{\vec{m}}_{\L,M,\l,a,\tau}  \;\; C^{\vec{e}}_{[N],\l,a,\check \tau}\;\;  w^{\vec{e}}_{\vec{m}}} \label{eq:defchiralop}
\end{equation}

The physics of eighth-BPS states and their partition functions from
both the field theory and the supergravity point of view (where they
are product of the half-BPS supergravity multiplet) were studied in
\cite{0510251}.  The quarter and eighth-BPS gauge invariant operators
should be related to giant gravitons generalizing the analogous
connection in the half-BPS case. It has been argued that the physics
of the eighth-BPS giants \cite{0010206} is given by the dynamics of
$N$ particles in a 3D simple harmonic oscillator
\cite{0207125,0507203,0606087,0606088}.  The states built from these
commuting bosons can be counted combinatorially in terms of vector
partitions \cite{0802.3015}.  Giant gravitons with strings attached
were considered in \cite{0604123,ggstringsattach}.  In the
supergravity approximation generalisations of the LLM solution
\cite{0409174} were investigated in
\cite{0606199,0704.2233,0802.0735}.

\subsection{Counting at finite $N$}

Counting the multiplicities of the operators we have constructed
\eqref{eq:defchiralop} for a given $G$ rep $\L$ gives
\begin{equation}
  \# \textrm{ of operators for } \L\;\; =\;\; \sum_{\l(S_n)} \mult(V_F^{\otimes n} \to \L\otimes\l)\;\;\mult((V_{\nat}^{S_N})^{\otimes n} \to [N]\otimes \l)
\end{equation}
Summing over all representations with their characters $\chi_\L({\bf
  x})$ gives the partition function for the chiral ring
\begin{equation}
  Z^{cr}_{U(N)}({\bf x}) =\sum_n \sum_{\L(G)} \sum_{\l(S_n)}  \mult(V_F^{\otimes n} \to \L\otimes\l)\;\;\mult((V_{\nat}^{S_N})^{\otimes n} \to [N]\otimes \l) \;\;\chi_\L({\bf x}) \label{eq:part}
\end{equation}
For global symmetry group $G=U(2)$ the Young diagram for the $S_n$ representation $\l$ is the
same as that for the $U(2)$ representation $\L$, so $\L=\l$ and
$\mult(V_{\bf 2}^{\otimes n} \to \L\otimes\l) = 1$.  This gives
\begin{equation}
  Z^{cr}_{U(N)}(x,y) =\sum_n \sum_{\L(U(2),S_n)}  \;\mult((V_{\nat}^{S_N})^{\otimes n} \to [N]\otimes \L) \;\;\chi_\L(x,y) \label{eq:part2}
\end{equation}
In the remainder of this section we verify the counting in the
partition function \eqref{eq:part2} by comparing it to known formulae
for the $U(2)$ sector, i.e. the genuine quarter-BPS operators.  The
generating function for $ Z^{cr}_{U(N)}(x,y)$ is given by
\cite{0510251}\cite{0608050}\cite{0701063}
\begin{equation}
  Z^{cr}(\n,x,y) = \prod_{n,m=0}^\infty \frac{1}{1-\n x^ny^m} = \sum_{N=0}^\infty \n^N  Z^{cr}_{U(N)}(x,y)
\end{equation}
In \cite{0704.1038} Dolan showed that
\begin{equation}
  Z^{cr}_{U(N)}(x,y) = \sum_{K \vdash N} s_K(1,x,x^2,\dots) s_K(1,y,y^2,\dots) \label{eq:Dolan}
\end{equation}
where the sum is over partitions $K$ of $N$. $s_K(1,q,q^2,\dots)$ is
the Schur polynomial defined for the partition $K$ by
\begin{equation}
    s_K(1,q,q^2,\dots)  = \frac{1}{N!} \sum_{\s \in S_N} \chi_K(\s)  \tr\left(\s \; Q^{\otimes N}\right)
\end{equation}
$Q$ is the infinite matrix with $(1,q,q^2,q^3,\dots)$ on its diagonal.

Note that our formula \eqref{eq:part2} is a refinement of
\eqref{eq:Dolan} in that it can count the chiral ring operators for
each representation of $U(2)$ rather than as a total sum.  A proof
that \eqref{eq:part2} and \eqref{eq:Dolan} are the same can be found
in Appendix Section \ref{sec:CRproof}.

\subsection{Check of counting for half-BPS operators}

In the half-BPS case the global symmetry group representation is
symmetrised $\L = \l = [n]$. The counting of the chiral ring gives
\begin{align}
  \mult((V_{\nat}^{S_N})^{\otimes n} \to [N]\otimes [n]) = p(n,N)
\end{align}
$p(n,N)$ is the number of partitions of $n$ into at most $N$ parts.  This
counts the Schur polynomials $\ket{R}$ discussed in Section
\ref{sec:schurpoly} where $R$, as a representation of $U(N)$, is a Young
diagram with $n$ boxes and at most $N$ rows.

The combinatorics of this counting can be connected directly with the
half-BPS supergravity geometries constructed by Lin, Lunin and
Maldacena \cite{0409174} (LLM) in the bulk.  In \cite{0111222,0403110}
it was shown that the half-BPS sector may be reduced to a complex
matrix model.  This in turn can be reduced to a system of the $N$
eigenvalues in a harmonic ocillator.  The eigenvalues become fermionic
due to the change in the path integral measure; their excitation
levels above the ground state then map to a partition into $N$ parts,
corresponding to the Young diagrams $R$ for the Schur polynomials.
The fermions can be represented as a Fermi droplet in phase space,
where a filled circle is the ground state and disturbances of this are
excitations.  The Young diagram $[1^N]$ gives each eigenvalue one
excitation, leaving a hole in the filled circle Fermi droplet (this is
the giant graviton expanded in the $S^5$).  $[N]$ gives only the top
eigenvalue a large excitation, leaving a small blob separated from the
filled Fermi droplet of the ground state (the giant graviton in the
$AdS_5$).

Approaching from the supergravity side, Lin, Lunin and Maldacena
\cite{0409174} (LLM) searched for all the half-BPS geometries with
$SO(4) \times SO(4) \times \R$ symmetry which are asymptotically
$AdS_5 \times S^5$.  They found smooth solutions determined by a
bi-coloured plane, which correspond exactly to the Fermi droplets of
the gauge theory matrix model.  Geometries with extremely large $R$
charge are similar to incipient black hole states and can be studied
as such \cite{0508023}.

\subsection{A combinatorial description of the quarter-BPS operators}

Given the connection between the LLM supergravity solutions and the
combinatorics of how half-BPS operators are counted for given global
charges, it is worth pursuing potential descriptions of the
quarter-BPS counting that might connect with quarter-BPS supergravity
solutions.  Investigations in the supergravity approximation for
quarter- and eighth-BPS geometries have been carried out in
\cite{0606199,0704.2233,0802.0735}.  Here we give a partial
description of the combinatorics from the gauge theory point of view.
The key result we show is that
\begin{equation}
  \boxed{\parbox{12cm}{ The number of quarter-BPS operators for the
      $U(2)$ representation $\L$ is the number of times the trivial
      $S_N$ representation $[N]$ appears when you decompose the
      representation of $U(N)$ with the same Young diagram $\L$ into
      irreps of its subgroup $S_N \subset U(N)$.}} \label{eq:14conj}
\end{equation}

From \eqref{eq:part2}, for a given representation $\L$ of $U(2)$ the
number of quarter-BPS operators is
\begin{equation}
  \mult((V_{\nat}^{S_N})^{\otimes n} \to [N]\otimes \L)
\end{equation}
This multiplicity can also be expressed in another way.  Starting from
\eqref{eq:SNSW} we could have treated $V_{\nat}^{S_N}$ as the
$N$-dimensional fundamental of $U(N)$, $V_{\bf N}^{U(N)}$, since they
are both of the same size.  If we had decomposed $n$ copies of this
representation using $U(N)$ and $S_n$ instead of $S_N$ and $S_n$ we
would have had a multiplicity-free decomposition
\begin{equation}
  \left(  V^{U(N)}_{\bf N} \right)^{\otimes n} = \bigoplus_{\k\vdash n} \;V_{\k}^{U(N)} \otimes V_{\k}^{S_n}
\end{equation}
Now identifying this with \eqref{eq:SNSW} we conclude that
$\mult((V_{\nat}^{S_N})^{\otimes n} \to K\otimes \k)$ is the number of
times $V_{K}^{S_N}$ appears when we decompose $V_{\k}^{U(N)}$ into
irreps of the subgroup $S_N \subset U(N)$
\begin{equation}
V_{\k}^{U(N)} = \bigoplus_{K\vdash N} \mult((V_{\nat}^{S_N})^{\otimes n} \to K\otimes \k)\;\;V_{K}^{S_N}
\end{equation}
Thus the number of quarter-BPS operators for the $U(2)$ representation
$\k=\L$ is the number of times the $S_N$ representation $[N]$ appears
when you decompose the representation $\k=\L$ of $U(N)$ into irreps of
its subgroup $S_N$.  In the remainder of this section we will try to
describe this number better and we will only partially succeed.

The states of $V_{\k}^{U(N)}$ consist of semi-standard tableaux where
we fill the $n$ boxes of the Young diagram $\k$ with numbers in $\{1,
\dots N\}$ such that they are weakly increasing along the rows and
strongly increasing down the columns.  Label the `field content' of
each semi-standard tableau by $\vec{\m}$ so that we have $\m_1$ 1's,
$\m_2$ 2's, \dots and $\m_N$ $N$'s.  $\sum_{i=1}^N \m_i = n$ so that $\vec{\m}$ is an ordered partition of $n$ into at most $N$ parts.

Under permutations in $S_N$ the $N$ fields transform as the natural
representation of $S_N$.  The overall (unordered) partition type of
each field content is invariant under $S_N$ so that for example for
$S_4 \subset U(4)$ the field content $\vec{\m} = (2,1,0,0)$ can be
transformed into $(1,0,0,2)$ but not into $(3,0,0,0)$:
\begin{equation}
  (142) \in S_4: \;\; \tyoung(11,2) \;\mapsto\; \tyoung(44,1) = -\,\tyoung(14,4)
\end{equation}
This means that we can partially reduce $V_{\k}^{U(N)}$ into
(reducible) representations $V_{\k,[\vec{\m}]}^{S_N}$ of $S_N$ 
corresponding to the equivalence classes of $\vec{\m}$.

For a given $[\vec{\m}]$ there are a certain number of fields which
don't appear in $\vec{\m}$.  Label this number of $0$'s in $\vec{\m}$
by $m_0$.  Since the semi-standard tableaux are left invariant by
permutations of these fields, $V_{[\vec{\m}]}^{S_N}$ must take the
form of an outer product\footnote{The symmetric group outer product
  denoted $\circ$ is the same as the tensor product for unitary group
  Young diagrams described by the Littlewood-Richardson coefficients.}
with the trivial representation $[m_0]$ of $S_{m_0}$
\begin{equation}
  V_{\k,[\vec{\m}]}^{S_N} = V_{\k,[\vec{\m}]'}^{S_{N-m_0}} \circ V_{[m_0]}^{S_{m_0}}
\end{equation}
where $[\vec{\m}]'$ has the $m_0$ 0's of $[\vec{\m}]$ removed.
To summarise the result so far
\begin{equation}
V_{\k}^{U(N)} = \sum_{[\vec{\m}]}\; V_{\k,[\vec{\m}]'}^{S_{N-m_0}} \circ V_{[m_0]}^{S_{m_0}}
\end{equation}
Decomposing $V_{\k,[\vec{\m}]'}^{S_{N-m_0}}$ is an open problem.  When
$N-m_0=n$ then $[\vec{\m}]'=[1^n]$ so that there are $n$ different
fields appearing only once each in the semi-standard tableau.  Under
these conditions the semi-standard tableau becomes standard
(where the numbers along each row must strictly increase).  This means
that $V_{\k,[1^n]}^{S_{n}} = V_{\k}^{S_n}$ since the standard
tableaux label the states of $\k$ of $S_n$.

As an example consider the decomposition of $\k=\tyng(2,1)$ of $U(N)$ into $S_N$ representations
\begin{align}
V_{\tyng(2,1)}^{U(N)} & =  V_{\tyng(2,1),[\vec{\m}]' =[1,1,1]}^{S_{3}} \circ V_{[N-3]}^{S_{N-3}} +  V_{\tyng(2,1),[\vec{\m}]' =[2,1]}^{S_{2}} \circ V_{[N-2]}^{S_{N-2}} \nn \\
& = \tyng(2,1) \circ [N-3]\; +\;\left( \tyng(2)\; +\; \tyng(1,1)\right) \circ [N-2]  \nn \\
& =  [N-3,2,1] + 2[N-2,1,1]+ 2[N-2,2]+3[N-1,1] + [N]
\end{align}
The single appearance of the $S_N$ trivial representation $[N]$
corresponds to the chiral ring $U(2)$ operator $\tr(\Phi_r)
\tr(\Phi^rX)$.  This result degrades appropriately for $n> N$, so for example for $N=2$
\begin{equation}
V_{\tyng(2,1)}^{U(2)}  =  \tyng(2)\; +\; \tyng(1,1)
\end{equation}
More examples of this type are available \cite{code}.

In the simplest case when the $U(N)$ representation is totally
symmetric $\k=[n]$ (corresponding as a $U(2)$ representation to the
half-BPS operators) we know the full answer:
$V_{[n],[\vec{\m}]'}^{S_{N-m_0}} = [m_1]\circ [m_2] \circ \cdots \circ
[m_n]$ where $m_i$ is the number of $i$'s in $\vec{\m}$ ($\sum_{i=0}^n
m_i = N$, $\sum_{i=0}^n im_i = n$).  Thus
\begin{equation}
  V_{[n]}^{U(N)} = \sum_{[\vec{\m}]}  [m_0] \circ [m_1] \circ \cdots \circ [m_n]
\end{equation}
The sum on the right is over symmetric group outer products that give
representations of $S_N$.  The trivial representation $[N]$ appears
once in this outer product for each $[\vec{\m}]$ so that it appears in
total $p(n,N)$ times, as known for the half-BPS case.

Finding such a graphical description of the decomposition of more
general representations of $U(N)$ into irreps of $S_N$ is an
outstanding problem that is also of interest to mathematicians
\cite{halverson}.  Solving it would give us a good combinatorial
description of the multiplicity of quarter-BPS operators for each
$U(2)$ representation $\L$.

\section{Conclusion}

The purpose of studying the combinatorics of the weakly-coupled gauge
theory is to understand better the putative weak-weak dual worldsheet
theory.  It is hoped that such a theory for the small radius $AdS_5
\times S^5$ background geometry is related by a weak-strong duality to
the standard large radius closed string theory.  Such relationships
are known for sigma models of compact symmetric superspaces
\cite{0809.1046,0908.0878}.

In Section \ref{sec:review} we reviewed the non-planar expansion of
extremal correlation functions for half-BPS operators.  The expansion
can be captured by the class algebra of the symmetric group, which
introduces a factor of $\frac{1}{N}$ each time a trace (i.e. a
permutation cycle) is cut or joined.  The combinatorics of the
non-planar corrections can be modelled by a two-dimensional theory on
higher genus surfaces with rings of spins connected by propagators
that do not cross \cite{0205033,0205089}.  As previously shown in
\cite{0205089,0206248,0209215}, in the BMN limit of operators with
many fields the correlation functions factorise into planar
three-point functions.

The half-BPS sector is especially simple because the fields in each
operator are symmetrised.  In Section \ref{sec:general} we showed how
representations of the global symmetry group are compatible with
specific trace structures.  This is important because we expect the
spectrum of the dual string theory to be organised into
representations of $PSU(2,2|4)$, while the trace structure should
roughly correspond to the multi-string Hilbert space.  We also showed
that much of the behaviour of the half-BPS sector is universal for all
operators: in the free theory the cut-and-join operators capture the
non-planar expansion and correlation functions of general states
factorise when the number of fields is large.

The weakly-coupled theory is discontinuously different to the
zero-coupling theory.  To find the correct state space our best guide
is to look for operators with well-defined conformal dimensions,
i.e. eigenstates of the dilatation operator.  After the half-BPS
operators, the next simplest sector is the $U(2)$ operators.  Using
technology from the free theory we showed in Section \ref{sec:weak}
how to construct the genuine quarter-BPS operators at weak coupling,
including all their $\frac{1}{N}$ corrections, by using their
orthogonality in the non-planar two-point function to the newly
descendant anomalous operators.  This fits the general expectation
that the $\frac{1}{N}$ expansion is not physical at non-zero coupling
(the correct stringy expansion is in $\frac{\l}{N}$) but $\frac{1}{N}$
corrections are important in identifying the right mapping between
SYM operators and states in the bulk.  In addition to
constructing the quarter-BPS operators we also have a novel
combinatorial way to count them in Section \ref{sec:weyl}, which may
be useful in finding quarter-BPS geometries analogous to the half-BPS
solutions constructed by LLM \cite{0409174}.

Understanding precisely how closed strings arise from $\cN=4$
super Yang-Mills is an on-going and challenging problem.  The
worldsheet theory must somehow capture the finite discrete
combinatorics of the gauge theory.  In the absence of a final model
for the dual of the weakly-coupled gauge theory, we hope analysis of
the field theory in this limit will shed some light on this problem.

\subsection{Further directions}

\begin{itemize}
\item Following the rudimentary comments in Section
  \ref{sec:worldsheet} a priority is to understand the dual discrete
  worldsheet theory better.  The conjecture \eqref{eq:gopconj}
  suggests connections between how homotopic propagators bunch in
  Gopakumar's model and the symmetric group conjugacy classes.  The
  universality of the factorisation suggests that the dual
  of the free theory is somewhat insensitive to the sector of the global
  symmetry group.
\item Part of the purpose of refocusing the finite $N$ technology on
  traces was to find the non-planar eigenstates of the dilatation
  operator.  We have done this for the simplest case by finding the
  $\frac{1}{N}$ corrections to the quarter-BPS operators; further
  results for anomalous operators are desirable. The $\frac{1}{N}$
  corrections to the anomalous eigenstates are not given by the
  conjugacy-class-invariant cut-and-join operators, see the discussion
  at the end of Appendix \ref{sec:42}.  However the dilatation
  operator \eqref{eq:roughdil} is not a conjugacy class invariant of
  $S_n$ either, so perhaps there is scope for a description of the
  anomalous eigenstates using cut-and-join operators that vary over
  each conjugacy class (while inheriting its $N$-dependence).
  Extending the three-string vertex \eqref{eq:messsy3pt} to one-loop
  eigenstates would be exciting.
\item There are several important features of $\cN=4$ SYM that this
  formalism doesn't incorporate.  As explained in Section
  \ref{sec:gengroup} we can do the Schur-Weyl duality for global
  groups $SO(2,4)$, $SO(6)$ and $U(K_1|K_2)$ but the full $PSU(2,2|4)$
  representations involve subtle shortening conditions.  Higher loop
  corrections also change the lengths of operators in the $SU(3|2)$
  sector \cite{0310252}, of which we haven't taken any account.
\item In this paper we have focused on extremal correlation functions.
  The full non-planar expansion of non-extremal three-point functions
  in the free theory was computed in \cite{thesis}. The structure was
  very similar so we expect many of the results here to follow
  through.
\item We have only solved the chiral ring counting problem as
  formulated in \eqref{eq:14conj} for special cases.  Finding a
  general graphical way of reducing representations of $U(N)$ to
  representations of its subgroup $S_N$ is an important mathematical
  problem in itself.
\item In this approach the underlying fields are replaced by
  gauge-invariant trace operators and interactions are replaced by the
  cutting and joining of traces.  Connections with the collective
  field theory approach to $\cN=4$ SYM \cite{collft} could be
  explored further.

\end{itemize}

\subsection{Acknowledgements}

We thank for many stimulating conversations Chong-sun Chu, Paul
Heslop, Antal Jevicki, Yusuke Kimura, Robert de Mello Koch, Shiraz
Minwalla, Jurgis Pasukonis, Sanjaye Ramgoolam, Rodolfo Russo, Volker Schomerus, Bill
Spence, Daniel Thompson and David Turton.  In particular we are
grateful to Robert de Mello Koch, Sanjaye Ramgoolam and Volker
Schomerus for reading drafts of the paper.

\begin{appendix}

\section{Genus $\geq 2$}\label{sec:highergenus}

\subsection{Genus 2}

The degree 4 cut-and-join operator is
\begin{equation}
  \frac{1}{N^4}\left( \S_{[5]}+ \S_{[4,2]}+\S_{[3,3]}+\S_{[3,2,2]}+\S_{[2,2,2,2]} \right)
\end{equation}
It gives the genus 2 contribution to the 2-point function
(cf. equation (2.6) of \cite{0205033})
\begin{equation}
  \corr{ \tr(X^{\dagger n}) \tr(X^n)}_{g=2} = nN^{n-4} \left[ 8 \binom{n}{5} +24\binom{n}{6} +12\binom{n}{6} +49\binom{n}{7}  +21\binom{n}{8}\right]
\end{equation}
corresponding to each splitting operator.  The final factor 21
comes from \eqref{eq:minw}. For $n$ large
\begin{equation}
   \corr{ \tr(X^{\dagger n}) \tr(X^n)}_{g=2} \to \frac{1}{4!}\bra{n} \S_{[2]}^4 \ket{n}
\end{equation}
To see how the factorisation works, insert schematic complete sets of
states (see equation \eqref{eq:S2onn1n2} for the exact form of the image of $\S_{[2]}$ on $\ket{n_1,n_2}$)
\begin{equation}
  \bra{n} \S_{[2]}\frac{\ket{n_1,n_2}\bra{n_1,n_2}}{\braket{n_1,n_2}{n_1,n_2}}\S_{[2]}\left( \frac{\ket{n}\bra{n}}{\braket{n}{n}} + \frac{\ket{n_1,n_2,n_3}\bra{n_1,n_2,n_3}}{\braket{n_1,n_2,n_3}{n_1,n_2,n_3}} \right) \S_{[2]}  \frac{\ket{n_1,n_2}\bra{n_1,n_2}}{\braket{n_1,n_2}{n_1,n_2}}\S_{[2]} \ket{n}
\end{equation}
This encodes $1\to 2 \to 1 \to 2\to 1$ and $1\to 2 \to 3 \to 2\to 1$.

\subsection{General genus}

We generally expect the form for the degree $k$ cut-and-join operator
\begin{equation}
  \frac{1}{N^k}\left( \S_{[k+1]}+ \S_{[k,2]}+\cdots + \S_{[3,2^{k-2}]}+\S_{[2^k]} \right)
\end{equation}
and for the 2-point function for $k=2g$
\begin{equation}
  nN^{n-k}\left[  c_{[k+1]}\binom{n}{k+1} + c_{[k,2]}\binom{n}{k+2} +\cdots + c_{[3,2^{k-2}]}\binom{n}{2k-1} +c_{[2^k]}\binom{n}{2k} \right]
\end{equation}
where $c_{[2^k]}$ is given by
\begin{equation}
  c_{[2^k]} = \frac{(4g)!}{2^{2g}(2g+1)!} = \frac{1\cdot 3\cdot 5 \cdots (4g-1)}{2g+1}  \label{eq:minw}
\end{equation}
as described in \cite{0205033,0205089}.  In the large $n$ limit
\begin{equation}
    \frac{(4g)!}{2^{2g}(2g+1)!} \binom{n}{2g} \sim   \frac{(4g)!}{2^{2g}(2g+1)!} \frac{n^{4g}}{(4g)!} = \frac{n^{4g}}{2^{2g}(2g+1)!}
\end{equation}
Summing over $g$ with $N^{-2g}$ we get $\frac{2\sinh(g_2/2)}{g_2}$ with
the BMN non-planar coupling $g_2 = \frac{n^2}{N}$ \cite{0205033,0205089}.

\section{Extremal $k$-point functions}\label{sec:4point}

For the extremal four-point function the action of
the vertex \eqref{eq:2vertex} on $\ket{n}$ in more detail is
\begin{align}
  \S_{[3]} \ket{n} &= \binom{n}{3} \ket{n} + \sum_{[n_1,n_2,n_3]} \frac{2nn_1n_2n_3}{|\Sym([n_1,n_2,n_3])|} \ket{n_1,n_2,n_3}\nn \\
  \S_{[2,2]} \ket{n} &= \binom{n}{4} \ket{n} +  \sum_{[n_1,n_2,n_3]} \frac{(n^2-3n)n_1n_2n_3}{|\Sym([n_1,n_2,n_3])|} \ket{n_1,n_2,n_3}
\end{align}
Once again the $\S_{[2,2]}$ term dominates over $\S_{[3]}$ as a
function of $n$, so the factorisation into two planar three-point
functions for large $n$ applies here too.

Applying the techniques used in Footnote \ref{eq:ijdetail} for
$\S_{[2]}$ it can be shown that when an element of $\S_{[3]}$ splits a
single trace into three the propagators are bunched into three groups
like the $Y$ diagram from \cite{freehigher}.  However, when an element
of $\S_{[2,2]}$ is applied the trace is split into four and then two
are rejoined out of order, just like the lollipop diagram of
\cite{freehigher}.

Adding the two terms together, the leading planar term of the extremal
four-point function is
\begin{equation}
  \frac{1}{N^2} \bra{n_1,n_2,n_3}\left( \S_{[3]}+\S_{[2,2]} \right)\ket{n} = \frac{n(n-1)n_1n_2n_3}{N^2}
\end{equation}
The general formula for the leading term of the extremal $k$-point function is
\begin{equation}
 \frac{1}{N^n}  \corr{ \tr(X^{\dagger n}) \tr(X^{n_1})\cdots \tr(X^{n_k})}_{\textrm{leading}}=  \frac{n!}{(n-k+1)!N^{k-2}} \prod_{i=1}^k n_i
\end{equation}

\section{Properties of $\O_n$}\label{sec:on}

$\O_n$ appeared in studies of 2d Yang-Mills
\cite{2dYM,9411210} when expanding unitary
group dimensions in $\frac{1}{N}$.  The relation to $U(N)$ dimensions
is given by
\begin{equation}
   \left( \dim R\right)^m  = \left(\frac{N^nd_R}{n!}\right)^m\frac{\chi_R(\O_n^m)}{d_R} \label{eq:Ondim}
\end{equation}
This equation can be reversed to give the expansion of $\O_n^m$ in
terms of dimensions
\begin{equation}
  \O_n^m =  \frac{1}{n!} \sum_{\s \in S_n} \sum_{R\in P(n)} d_R \left(\frac{n! \dim R}{N^nd_R}\right)^m\; \chi_R(\s)\;\s \label{eq:OfromDim}
\end{equation}
$\O_n^{-1}$ only exists if $N \geq n$. Its expansion is an infinite
series in $\frac{1}{N}$
\begin{equation*}
  \O_n^{-1} = 1 -\frac{1}{N} \S_{[2]} + \frac{1}{N^2} \left[\binom{n}{2}  + 2\S_{[3]} + \S_{[2,2]}  \right] - \frac{1}{N^3}\left[\frac{n^2+3n-8}{2}\S_{[2]} + 5 \S_{[4]}+ 2\S_{[3,2]} + \S_{[2,2,2]}\right]\cdots
\end{equation*}
To calculate the contribution for each $\S_C$ we can use
\eqref{eq:OfromDim} to get a finite series.  For example
\begin{align}
  \O_2^{-1} & = \frac{N^2}{N^2-1} \left(1 -\frac{1}{N} \S_{[2]}  \right) \nn \\
  \O_4^{-1}&  = \frac{N^6}{(N^2-1)(N^2-4)(N^2-9)} \left( 1 -\frac{1}{N}\S_{[2]} + \frac{1}{N^2} \left[-8+ 2\S_{[3]} + \S_{[2,2]}  \right] + \frac{1}{N^3}\left[4\S_{[2]} - 5 \S_{[4]}\right]\right.\nn \\
& \left.\hspace{6cm} + \frac{1}{N^4}\left[6-3\S_{[3]} + 6 \S_{[2,2]}\right] \right)
\end{align}
The exact exponential of $\O_n$ (cf. equation (2.5) of \cite{9407176})
is
\begin{equation}
  \O_n = \exp\left(\frac{1}{N} \S_{[2]} - \frac{1}{2N^2}\left[\binom{n}{2} + \S_{[3]}\right] + \frac{1}{3N^3} \left[(2n-3) \S_{[2]} + \S_{[4]}\right] + \cO\left(\frac{1}{N^4}\right) \right)\label{eq:exactexp}
\end{equation}
Note that the additional terms that are exponentiated are all
subleading in $\frac{n^2}{N}$ (which is the same as the BMN expansion
parameter $g_2 = \frac{J^2}{N}$ in the half-BPS sector when
$n=\Delta=J$).

\section{Basis details}

Symmetric group identities used here are listed in Appendix B of
\cite{thesis}.

\subsection{Inverting}\label{sec:invert}

We fix the permutation $\a$ in its conjugacy class.  If $\a =
[n_1,n_2, \dots n_k]$ then $\a = (1 \cdots n_1)(n_1+1 \cdots
n_1+n_2) \cdots ( \cdots n)$.  Any trace operator with field content and trace structure $\a$ can be written
\begin{equation}
  \tr(\a\r {\bf X}^{\vec{\m}} \r^{-1})
\end{equation}
for some $\r \in S_n$.  To get this trace from the $\ket{\L,M;\a,\g}$ basis
\begin{align}
  \sum_{\L,\b,\g} d_\L D^{\L}_{pq}(\rho) S_{p\g}^\a B_{q\b}^{\vec{\m}} \ket{\L,{\vec{\m}},\b;\a,\g}  
& = \sum_\L \frac{1}{n!} \sum_{\s \in S_n} d_\L \chi_\L(\s \rho^{-1}) \tr(\a\s {\bf X}^{\vec{\m}} \s^{-1}) \nn \\
& =  \tr(\a\r {\bf X}^{\vec{\m}} \r^{-1}) \label{eq:inversion}
\end{align}
So the basis $\ket{\L,{\vec{\m}},\b;\a,\g}$ does hit all the operators.

\subsection{Counting}\label{sec:counting}

The counting formula \eqref{eq:tracecount} exactly matches known
counting formulae.  For example, counting single trace operators $\a =
(123 \cdots n)$ where $\Sym(\a) = \Z_n$, we can read off the
coefficients from equation (3.6) of \cite{0405057} where they're
counting the appearance of YT-pletons for $hs(2,2|4)$ in the single
trace partition function.

All multi-trace operators with a given symmetrisation $\L$ are counted by
\begin{align}
\sum_{\a\in P(n)} S(\a,\L) = &  \sum_{\a \in P(n)}  \frac{1}{|\Sym(\a)|} \sum_{\rho \in \Sym(\a)} \chi_{\L}(\rho) \nn \\
=& \frac{1}{n!} \sum_{\a \in S_n} \sum_{\rho \in S_n} \delta(\rho\a\r^{-1}\a^{-1}=1)\;\; \chi_{\L}(\rho)
\end{align}
$\rho \in |\Sym(\a)|$ is enforced using a $\delta$-function.
Now expand the $\delta$-function using a sum over $S_n$
representations $R$
\begin{align}
\sum_{\a\in P(n)} S(\a,\L)
=& \frac{1}{n!} \sum_{\a \in S_n} \sum_{\rho \in S_n} \frac{1}{n!}\sum_{R\in P(n)} d_R\; \chi_R(\rho\a\r^{-1}\a^{-1})\;\; \chi_{\L}(\rho) \nn \\
=& \frac{1}{n!} \sum_{\a \in S_n} \sum_{\rho \in S_n} \frac{1}{n!}\sum_{R \in P(n)} \chi_R(\rho)\;\chi_R(\a\r^{-1}\a^{-1})\;\; \chi_{\L}(\rho) \nn \\
=&\sum_{R \in P(n)}\; \frac{1}{n!} \sum_{\rho \in S_n}\chi_R(\rho)\;\chi_R(\r)\; \chi_{\L}(\rho) 
\end{align}
This matches the large $N$ result from \cite{0704.1038,0711.0176}.

\subsection{Three-point vertex}\label{sec:3ptvertex}

We want to provide the details for equation \eqref{eq:messsy3pt}.  In
particular we want to find the decomposition coefficients between the
double-trace $[n_1,n_2]$ and the product of two single-traces
\begin{align*}
 \ket{\L,M;\a=[n_1,n_2],\g} = \sum_{\L_i,M_i,\g_i} C^{\L,M,\g}_{\{\L_i,M_i,\g_i\}}\;\; \ket{\L_1,M_1;\a_1=[n_1],\g_1}\otimes  \ket{\L_2,M_2;\a_1=[n_2],\g_2} 
\end{align*}
A vanilla double-trace operator defined by $\a=[n_1,n_2]$ and $\s$ can
be written as the product of two traces $\a_1=[n_1]$ and
$\a_2=[n_2]$ with $\vec{\m}_1$, $\vec{\m_2}$, $\s_1$ and $\s_2$
depending on how the original double-trace laced the fields
\begin{equation}
  \tr(\s^{-1}[n_1,n_2]\s \;{\bf X}^{\vec{\mu}} ) = \tr(\s_1^{-1}[n_1]\s_1\; {\bf X}^{\vec{\mu}_1} )\;\tr(\s_2^{-1}[n_2]\s_2\; {\bf X}^{\vec{\mu}_2} )\label{eq:splitdoule}
\end{equation}
To get the coefficients $C^{\L,M,\g}_{\{\L_i,M_i,\g_i\}}$ use the
definition of $ \ket{\L,M;\a=[n_1,n_2],\g}$ in terms of
$\tr(\s^{-1}\a\s \;{\bf X}^{\vec{\mu}} )$ from \eqref{tracegambas} and
then use \eqref{eq:splitdoule} and the inversion procedure
\eqref{eq:inversion} on the $\tr(\s_i^{-1}[n_i]\s_i\; {\bf
  X}^{\vec{\mu}_i})$.

\subsection{Change of basis to finite $N$ basis}\label{sec:gamtotau}

We want to describe the matrix of the change of basis to
\begin{align}
     \cO[ \L, \mu ,\b; R, \tau]= \frac{1}{n!}\sum_{\rho}
     B^{\vec{\m}}_{j \b} \; S^{\tau ,}{}^{ \L }_{ j }\;{}^{R}_{p
     }\;{}^{R}_{q}\;\; D_{pq}^R(\rho)\;\; \tr(\rho\; {\bf X}^{\mu} )
\end{align}

A first step to fix the canonical members of each conjugacy class and
sum over their conjugations
\begin{equation}
  \sum_{\rho \in S_n} f(\rho) = \sum_{\a \in P(n)}  \frac{1}{|\Sym(\a)|} \sum_{\tau \in S_n} f(\tau^{-1} \a \tau)
\end{equation}

We get
\begin{align}
     \cO[ \L, \mu ,\b; R, \tau]= \sum_{\a \in P(n)} \frac{1}{|\Sym(\a)|} \sum_{\g} S^\a_{j \g} \; S^{\tau  ,}{}^{  \L }_{  j
  }\;{}^{R}_{p }\;{}^{R}_{q}\;\; D_{pq}^R(\a)\;\;\cO[\L,\m,\b;\a,\g]
\end{align}

The other way round
\begin{align}
\cO[\L,\m,\b;\a,\g]  = \sum_{R,\tau}\frac{d_R}{d_\L} S^{\a}_{j\g}\; S^{\tau  ,}{}^{  \L }_{  j
  }\;{}^{R}_{p }\;{}^{R}_{q}\;\; D_{pq}^R(\a)\;\;\cO[ \L, \mu ,\b; R, \tau]
\end{align}
These formulae are compatible with each other.

\section{Trace operator examples}\label{sec:trexamples}

Computer code used for these examples is available on the internet \cite{code}.

\subsection{$\L=[2,2]$ example}\label{sec:22}

The operators for $\L=[2,2]$ are listed in Table \ref{tab:kupto3}.
This case is not so interesting because there is never any
multiplicity for the $\a$.
\begin{table}[h!]
  \centering
  \begin{tabular}[h]{c | c | c }
$\a$&$S(\a,\L=[2,2])$& operator(s) \\
\hline
$[1,1,1,1]$ & 0 & \\
$[2,1,1]$ &1 &   $\cO_1 = \tr(\Phi^r\Phi^s) \tr(\Phi_r)\tr(\Phi_s)$\\
$[2,2]$ & 1 &   $\cO_2 = \tr(\Phi^r\Phi^s) \tr(\Phi_r\Phi_s)$\\
$[3,1]$ & 0 & \\
$[4]$ & 1 & $\cO_3 = \tr(\Phi^r\Phi_r \Phi^s \Phi_s)$\\
  \end{tabular}
  \caption{Operators for $\L=[2,2]$.}
  \label{tab:kupto3}
\end{table}

For $\a = (12)$ decompose the projector \eqref{eq:tracedeomcp}
to get the $S^\a_{a}$
\begin{align}
  \frac{1}{|\Sym(\a)|} \sum_{\rho \in \Sym(\a)}D^\L_{ab}(\rho)&  = \frac{1}{4}\left(
  \begin{array}{cc}
    4  & 0 \\
    0 & 0
  \end{array}
  \right)  = \left(
  \begin{array}{c}
     1 \\
    0
  \end{array}
  \right)(1,0)  = S^{[2,1,1]}_{a}S^{[2,1,1]}_{b}
\end{align}
For $\a = (12)(34)$
\begin{align}
  \frac{1}{|\Sym(\a)|} \sum_{\rho \in \Sym(\a)}D^\L_{ab}(\rho)&  = \frac{1}{8}\left(
  \begin{array}{cc}
    8 & 0 \\
    0 & 0
  \end{array}
  \right)  = \left(
  \begin{array}{c}
     1 \\
    0
  \end{array}
  \right)(1,0) = S^{[2,2]}_{a}S^{[2,2]}_{b}
\end{align}
For $\a = (1234)$
\begin{align}
  \frac{1}{|\Sym(\a)|} \sum_{\rho \in \Sym(\a)}D^\L_{ab}(\rho)&  = \frac{1}{4}\left(
  \begin{array}{cc}
    1  & -\sqrt{3} \\
     -\sqrt{3} & 3
  \end{array}
  \right)  = \left(
  \begin{array}{c}
    -\tfrac{1}{2} \\
    \tfrac{\sqrt{3}}{2} 
  \end{array}
  \right)(-\tfrac{1}{2},\tfrac{\sqrt{3}}{2})= S^{[4]}_{a}S^{[4]}_{b}
\end{align}

We get
\begin{align}
  \ket{\L=[2,2],M=[2,2];\a=[2,1,1]} & = \frac{1}{6} \cO_1 \nn \\
  \ket{\L=[2,2],M=[2,2];\a=[2,2]} & = \frac{1}{6}  \cO_2 \nn \\
  \ket{\L=[2,2],M=[2,2];\a=[4]} & = \frac{7}{24}  \cO_3
\end{align}

At weak coupling the eigenstates of the dilatation operator are
\begin{align}
  & \D\; \cO_1 = 0 \nn \\
  & \D\; \left[\cO_2+ \frac{2}{N}\cO_3 \right] = 0 \nn \\
  & \D\; \cO_3  = -6N\; \cO_3  \label{eq:22eigen}
\end{align}
The only descendant operator is $\a^d = [4]$ corresponding to $\cO_3$.
The remainder $\a^c_1 = [2,1,1]$ and $\a^c_{2}=[2,2]$ give the leading
terms of the two genuine quarter-BPS operators following equation \eqref{eq:quarterdef}
\begin{align}
  \ket{\L=[2,2],M=[2,2];\a^c_1=[2,1,1];\perp} & = \O_4^{-1} \ket{\L=[2,2],M=[2,2];\a^c_1=[2,1,1]}\nn \\
& =\frac{N^4}{6(N^2-1)(N^2-4)} \left[\cO_1-\frac{1}{N}\cO_2 - \frac{2}{N^2}\cO_1  - \frac{2}{N^2}\cO_3 \right]\nn \\
  \ket{\L=[2,2],M=[2,2];\a^c_2=[2,2];\perp} & = \O_4^{-1} \ket{\L=[2,2],M=[2,2];\a^c_2=[2,2]}\nn \\
& = \frac{N^4}{6(N^2-1)(N^2-4)} \left[ \cO_2+ \frac{2}{N} \cO_3 -\frac{2}{N} \cO_1  \right]
\end{align}
These are a linear combination of the protected operators identified
in \eqref{eq:22eigen}.

\subsection{$\L=[4,2],\a = [4,2]$ example}\label{sec:42}

With trace structure $\a = [4,2] = (1234)(56)$ there are two
independent operators
\begin{align}
  \cO_1 & = \tr([X,Y][X,Y])\tr(XX) = \tr(\Phi^r\Phi_r \Phi^s \Phi_s) \tr(XX)\nn \\
& = 2\tr(XYXY)\tr(XX) - 2\tr(XXYY)\tr(XX) \nn \\
& \nn \\
  \cO_2 & = \tr(\Phi^r\Phi^s XX) \tr(\Phi_r \Phi_s)\nn \\
& = \tr(XXXX)\tr(YY) - 2\tr(XXXY)\tr(XY) + \tr(YYXX)\tr(XX)
\end{align}

Decompose the projector \eqref{eq:tracedeomcp} to get the two orthogonal $S^\a_{a\g}$
\begin{align}
  \frac{1}{|\Sym(\a)|} \sum_{\rho \in \Sym(\a)}D^\L_{ab}(\rho)&  = \frac{1}{8}\left(
  \begin{array}{cccc}
    8 & \cdots &  &  \\
    \vdots & \ddots & & \\
& & 2 & -2\sqrt{3} \\
&& -2\sqrt{3} & 6
  \end{array}
  \right) \nn \\
& = \left(
  \begin{array}{c}
    1 \\
    0 \\
    \vdots
  \end{array}
  \right)(1, 0, \cdots) + \left(
  \begin{array}{c}
    0 \\
    \vdots \\
    -\tfrac{1}{2}\\
    \tfrac{\sqrt{3}}{2}
  \end{array}
  \right)(0, \cdots,    -\tfrac{1}{2},
    \tfrac{\sqrt{3}}{2}) \nn \\
& = S^{[4,2]}_{a1}S^{[4,2]}_{b1} + S^{[4,2]}_{a2}S^{[4,2]}_{b2}
\end{align}
Note that the trace is 2 which is the correct number of operators for
this trace structure.

We get
\begin{align}
  \ket{\L=[4,2],M=[4,2];\a=[4,2],\g=1} & = \frac{1}{15}\left(\cO_2 + \frac{1}{6} \cO_1\right) \nn \\
  \ket{\L=[4,2],M=[4,2];\a=[4,2],\g=2} & = \frac{1}{9\sqrt{5}}  \cO_1
\end{align}
Fortunately we don't have to do any rearrangement to get the
descendant operator $\cO_1$ with commutators inside the trace:
$\{\a^d,\g^d\}=\{[4,2],2\}$ is precisely $\cO_1$.
$\{\a^c,\g^c\}=\{[4,2],1\}$ is orthogonal to $\{[4,2],2\}$ in the
planar inner product by construction so it is exactly the correct
operator for the leading term of the quarter-BPS state.  If a single symmetrised trace is written $A^{r_1 \cdots r_p} = \tr(\Phi^{(r_1} \cdots \Phi^{r_p)})$ then this operator is $\e_{r_1s_1} \e_{r_2s_2}A^{r_1r_211} A^{s_1s_2}$
 cf. equation (2.28) of
\cite{0301104}.\footnote{One might wonder why the answer dropped out
  so easily in this case.  We expect the leading terms of the
  eigenstates of $\Delta$ to be orthogonal in the planar tree-level
  two-point function.  Thus it's possible that by diagonalising the
  planar two-point function with the $\{\a,\g\}$ we hit on the same
  operators that are eigenstates of $\D$ to planar order. However this
  is not guaranteed and is most likely a coincidence in this case.}
\begin{align}
 & \ket{[4,2],[4,2];\a^c=[4,2],\g^c=1;\perp}\nn\\
 & = \O_6^{-1} \ket{[4,2],[4,2];\a^c=[4,2],\g^c=1}\nn \\
& = \frac{1}{15} \left[\cO_2+\frac{1}{6}\cO_1 + \frac{8}{3N}\tr(\Phi^r\Phi_r \Phi^s \Phi_sXX) -\frac{16}{3N} \tr(\Phi^r \Phi^s\Phi_r \Phi_sXX)\right. \nn \\
&- \frac{4}{3N} \tr(\Phi^r \Phi^s)\tr(\Phi_r \Phi_s)\tr(XX) - \frac{1}{N} \tr(\Phi^r \Phi^sXX)\tr(\Phi_r)\tr( \Phi_s) - \frac{1}{6N} \tr(\Phi^r\Phi_r \Phi^s \Phi_s)\tr(X)\tr(X) \nn \\
&\left.  -\frac{4}{N} \tr(\Phi^r \Phi^sX)\tr(\Phi_r \Phi_s)\tr(X)       +\frac{2}{N} \tr(\Phi^r \Phi^sX)\tr(\Phi_r X)\tr(\Phi_s)      +\cO\left(\frac{1}{N^2}\right) \right]
\end{align}
To show that this is correct to this order in $\frac{1}{N}$ we give
the leading terms of the relevant eigenstates of $\D$ for $\L=[4,2]$
(for which there are 15 independent trace operators in total) in Table
\ref{tab:42estates}.
\begin{table}[h!]
  \centering
  \begin{tabular}[h]{l | l }
 leading terms of eigenstates & eigenvalue \\
\hline
  $\tr(\Phi^r\Phi_r \Phi^s \Phi_sXX) + (-2 \pm \sqrt{5})\tr(\Phi^r\Phi^s \Phi_r  \Phi_sXX) + \cO\left(\frac{1}{N} \right)$ & $-(5\pm\sqrt{5})N +\cO(N^{-1})$ \nn \\
$\cO_1 -\frac{14}{N}\tr(\Phi^r\Phi_r \Phi^s \Phi_sXX) - \frac{2}{N} \tr(\Phi^r\Phi^s \Phi_r  \Phi_sXX) +\cO\left(\frac{1}{N^2}\right)  $   & $-6N +\cO(N^{-1})$ \nn \\
  $\cO_2 + \frac{1}{6} \cO_1+ \frac{8}{3N} \tr(\Phi^r\Phi_r \Phi^s \Phi_sXX) - \frac{16}{3N}\tr(\Phi^r\Phi^s \Phi_r  \Phi_sXX) +\cO\left(\frac{1}{N^2}\right)  $   & $0$ \nn \\
  $ \tr(\Phi^r \Phi^s)\tr(\Phi_r \Phi_s)\tr(XX) +\frac{2}{N} \cO_1 -\frac{8}{N} \cO_2 +\cO\left(\frac{1}{N^2}\right)  $   & $0$ \nn\\
$\tr(\Phi^r\Phi_r \Phi^s \Phi_s)\tr(X)\tr(X)$ & $-6N$ \nn \\
$ \tr(\Phi^r \Phi^sXX)\tr(\Phi_r)\tr( \Phi_s) + \frac{1}{6}\tr(\Phi^r\Phi_r \Phi^s \Phi_s)\tr(X)\tr(X)$ & $0$ \nn \\
 $\tr(\Phi^r \Phi^sX)\tr(\Phi_r \Phi_s)\tr(X)  +\cO\left(\frac{1}{N}\right)  $   & $0$ \nn \\
 $\tr(\Phi^r \Phi^sX)\tr(\Phi_r X)\tr(\Phi_s)    +\cO\left(\frac{1}{N}\right)  $   & $0$ 
  \end{tabular}
  \caption{Leading terms of some of the eigenstates of $\D$ for $\L=[4,2]$.}
  \label{tab:42estates}
\end{table}
Unfortunately the subleading terms in $\frac{1}{N}$ of the anomalous
eigenstates don't seem to be given in any obvious way by the
cut-and-join operators.  For example applying $\S_{[2]}$ to $\cO_1$ we get
\begin{align}
  \Sigma_{[2]} \cO_1 = & 8\tr(\Phi^r\Phi_r \Phi^s \Phi_sXX)  -4\tr(\Phi^r\Phi^s \Phi_r  \Phi_sXX) \nn \\&\quad - \tr(\Phi^r \Phi^s)\tr(\Phi_r \Phi_s)\tr(XX) +\tr(\Phi^r\Phi_r \Phi^s \Phi_s)\tr(X)\tr(X)
\end{align}
which doesn't match the correct $\frac{1}{N}$ term for the operator
leading with $\cO_1$ in Table \ref{tab:42estates}.

\section{Equivalence of one-loop formulations}\label{sec:1lpequiv}

Here we demonstrate the equivalence of the two one-loop actions of
\cite{0404066} and \cite{0801.2094} used in Section \ref{sec:further}.
Suppose $P_{ij}$ projects onto two sites $1 \leq i,j\leq n$.  If $i$
and $j$ sit within a single trace $\a = (i\,s_1\,j\,s_2)$ where
$s_1,s_2$ are some sequences of integers then
\begin{align}
  (i\,n\pd1)\,\a\,(j\,n\pd1) = (i\,s_1\,j)(n\pd1\,s_2) = \left\{ 
  \begin{array}{l}
    (i\,s_1\,j)(s_2) \quad \quad \textrm{if  } s_2 \neq 0 \\
    N(i\,s_1\,j) \quad \quad \textrm{  if  } s_2 = 0 
  \end{array}
  \right.
\end{align}
If we split off the first integer of $s_2=\a(j)\,s_2{}'$ then
\begin{equation}
  (i\,n\pd1)\,\a\,(j\,n\pd1) = (i\,\a(j))\,\a = (i\,s_1\,j)(\a(j)\,s_2{}') = (i\,s_1\,j)(s_2)
\end{equation}
which agrees if $s_2 \neq 0$.  If $s_2$ vanishes then $\a(j) = i$ and
$(i\,\a(j)) = N$, cf. \cite{0404066} equation (2.15).

If $i$ and $j$ sit within two different traces $\a= (i\,s_1)(j\,s_2)$
then
\begin{equation}
  (i\,n\pd1)\a(j\,n\pd1) = (j\,i\,s_1\,s_2) = (i\,\a(j))\,\a
\end{equation}

\section{Proof of chiral ring counting formula}\label{sec:CRproof}

 To show
that our result \eqref{eq:part} is the same as that in the literature
\eqref{eq:Dolan}, first note that for representations of $U(2)$ the
Young diagram for the $S_n$ representation $\l$ is the same as that
for the $U(2)$ representation $\L$, so $\L=\l$ and $\mult(V_{\bf
  2}^{\otimes n} \to \L\otimes\l) = 1$.

In general the multiplicity $\mult((V_{\nat}^{S_N})^{\otimes n} \to
K\otimes \k)$ in \eqref{eq:SNSW} can be calculated using projectors for $K$ and $\k$ in $(V_{\nat}^{S_N})^{\otimes n}$
\begin{equation}
\mult((V_{\nat}^{S_N})^{\otimes n} \to
K\otimes \k) =  \frac{1}{N!} \sum_{\sigma \in S_N} \chi_K(\s)  \frac{1}{n!} \sum_{\tau \in S_n} \chi_\k(\tau) \prod_i (\chi_{\nat} (\s^i))^{c_i(\tau)}\label{eq:multexpan}
\end{equation}
where $c_i(\tau)$ is the number of cycles of length $i$ in $\tau\in
S_n$.

For the specialisation to $\k = [n]$ we will also use
\begin{equation}
\mult((V_{\nat}^{S_N})^{\otimes n} \to
K\otimes [n]) = \textrm{coefficient of } q^n \textrm{ in } s_K(1,q,q^2,\dots)
\end{equation}
Alternatively this can be stated
\begin{equation}
  s_K(1,q,q^2,\dots)  =  \sum_{n=0}^\infty \mult((V_{\nat}^{S_N})^{\otimes n} \to
K\otimes [n])  \;\; q^n \label{eq:alstat}
\end{equation}

Our expression for the $U(2)$ partition function \eqref{eq:part2} for
$\tfrac{1}{4}$-BPS chiral ring states is
\begin{align}
  Z^{cr}_{U(N)}(x,y) & = \sum_\L  \mult((V_{\nat}^{S_N})^{\otimes n} \to
[N]\otimes \L)\;\;\chi_\L(x,y) \nn \\
   & = \sum_\L  \mult((V_{\nat}^{S_N})^{\otimes n} \to
[N]\otimes \L) \;\;\sum_{\m,\n} g([\m],[\n];\L)\; x^\m y^\n
\end{align}
where we've expanded out the Schur polynomial using the
Littlewood-Richardson coefficient $g$.  Next use \eqref{eq:multexpan}
and the formula for the Littlewood-Richardson coefficient $g$ from
\eqref{eq:kostka} to get
\begin{align}
  Z^{cr}_{U(N)}(x,y) & = \sum_\L  \frac{1}{N!} \sum_{\sigma \in S_N}   \frac{1}{n!} \sum_{\tau \in S_n} \chi_\L(\tau) \prod_i (\chi_{\nat} (\s^i))^{c_i(\tau)}   \;\;\sum_{\m,\n} \frac{1}{\m!\n!} \sum_{\rho \in S_\m \times S_\n}\chi_\L(\rho) \; x^\m y^\n  \nn \\
& = \sum_{\m,\n} \frac{1}{\m!\n!} \sum_{\rho \in S_\m \times S_\n}\;  \frac{1}{N!} \sum_{\sigma \in S_N}   \prod_i (\chi_{\nat} (\s^i))^{c_i(\rho)}  \; x^\m y^\n  \label{eq:toagree}
\end{align}

Next, working from Dolan's formula we use \eqref{eq:alstat} to get
\begin{align}
   Z^{cr}_{U(N)}(x,y) & = \sum_{K \vdash N} s_K(1,x,x^2,\dots) s_K(1,y,y^2,\dots) \nn \\
 & = \sum_{K \vdash N} \sum_{\m,\n} \mult((V_{\nat}^{S_N})^{\otimes \m} \to
K\otimes [\m]) \;\;\mult((V_{\nat}^{S_N})^{\otimes \n} \to
K\otimes [\n]) \;\;  x^\m y^\n
\end{align}
Now use \eqref{eq:multexpan} to get
\begin{align}
   Z^{cr}_{U(N)}(x,y)
 =& \sum_{K \vdash N} \sum_{\m,\n} \frac{1}{N!} \sum_{\s_1 \in S_N} \chi_K(\s_1)  \frac{1}{\m!} \sum_{\rho_1 \in S_\m}  \prod_i (\chi_{\nat} (\s_1^i))^{c_i(\rho_1)} \nn \\
& \frac{1}{N!} \sum_{\s_2 \in S_N} \chi_K(\s_2)  \frac{1}{\n!} \sum_{\rho_2 \in S_\n}  \prod_j (\chi_{\nat} (\s_2^j))^{c_j(\rho_2)} \;\;  x^\m y^\n \nn \\
  =& \sum_{\m,\n} \frac{1}{N!} \sum_{\s \in S_N}  \frac{1}{\m!} \sum_{\rho_1 \in S_\m}  \prod_i (\chi_{\nat} (\s^i))^{c_i(\rho_1)}   \frac{1}{\n!} \sum_{\rho_2 \in S_\n}  \prod_j (\chi_{\nat} (\s^j))^{c_j(\rho_2)} \;\;  x^\m y^\n\nn \\
 =& \sum_{\m,\n} \frac{1}{\m!\n!} \sum_{\rho \in S_\m \times S_\n}\;  \frac{1}{N!} \sum_{\sigma \in S_N}   \prod_i (\chi_{\nat} (\s^i))^{c_i(\rho)}  \; x^\m y^\n  
\end{align}
This is identical to \eqref{eq:toagree} so we are done.

\end{appendix}

\end{document}